\documentclass[useAMS,usenatbib,fleqn]{mnras}

\usepackage[utf8]{inputenc}
\usepackage{graphicx,color}
\usepackage{mathtools}	
\usepackage{amsmath}
\usepackage{amssymb}
\usepackage{pstricks} 

  \usepackage[british]{babel}             
  \usepackage{newtxtext}                  
  \usepackage[slantedGreek]{newtxmath}    
  \usepackage[T1]{fontenc}                

\def\apj{ApJ}

\def\mnras{MNRAS}
\def\aap{A\&A}
\def\apjl{ApJ}

\def\physrep{PhR}

\def\apjs{ApJS}
\def\pasa{PASA}

\def\pasp{PASP}

\def\araa{ARA\&A}

\def\aapr{ARA\&A}

\newcommand{\f}{_\mathrm{0}}					   	
\newcommand{\ma}{_\mathrm{max}}			   		

\newcommand{\lambdatilde}{\widetilde{\lambda}}
\newcommand{\lamtilde}{\lambdatilde}

\newcommand{\cut}{_\mathrm{c}}

\newcommand{\dynam}{_\mathrm{dyn}}

\newcommand{\oneenv}{_\mathrm{1,env}}

\newcommand{\soft}{_\mathrm{s}}
\newcommand{\softf}{_\mathrm{s,0}}

\newcommand{\inone}{_\mathrm{in,1}}
\newcommand{\intwo}{_\mathrm{in,2}}
\newcommand{\Kep}{_\mathrm{K}}
\newcommand{\Gn}{\mathrm{G}}
\newcommand{\init}{_\mathrm{i}}
\newcommand{\acc}{_\mathrm{acc}}

\newcommand{\proton}{_\mathrm{p}}
\newcommand{\Rsol}{\,R_\odot}
\newcommand{\Msol}{\,M_\odot}
\newcommand{\base}{_\mathrm{base}}
\newcommand{\amb}{_\mathrm{amb}}

\newcommand{\starmax}{_{*,\mathrm{max}}}

\newcommand{\cm}{\,{\rm cm}}

\newcommand{\g}{\,{\rm g}}
\newcommand{\gcmcmcm}{\,{\rm g\,cm^{-3}}}

\newcommand{\yr}{\,{\rm yr}}     
\newcommand{\da}{\,{\rm d}}     
\newcommand{\dynecmcm}{\,{\rm dyn\,cm^{-2}}}     
  
\title[Accretion in common envelope evolution]
{Accretion in common envelope evolution}
\author[L.\ Chamandy et al.]{Luke Chamandy$^{1}$\thanks{lchamandy@pas.rochester.edu},
Adam Frank$^{1}$\thanks{afrank@pas.rochester.edu}, 
Eric G.~Blackman$^{1}$\thanks{blackman@pas.rochester.edu},
Jonathan Carroll-Nellenback$^{1}$,
\newauthor 
Baowei Liu$^{1}$
, Yisheng Tu$^{1}$
, Jason Nordhaus$^{2,3}$
, Zhuo Chen$^{1}$
and Bo Peng$^{1}$\\
\\
$^{1}$Department of Physics and Astronomy, University of Rochester, Rochester NY 14618, USA\\
$^{2}$National Technical Institute for the Deaf, Rochester Institute of Technology, NY 14623, USA\\
$^{3}$Center for Computational Relativity and Gravitation, Rochester Institute of Technology, NY 14623, USA
}

\begin{document}


\maketitle

\begin{abstract}
Common envelope evolution (CEE) is presently a poorly understood, yet critical, process in binary stellar evolution.  
Characterizing the full 3D dynamics of CEE is difficult in part because simulating CEE is so computationally demanding.  
Numerical studies have yet to conclusively determine how the envelope ejects 
and a tight binary results, if only the binary potential energy is used  to propel the envelope.  
Additional power sources might be necessary and  accretion onto the inspiraling  companion is one such  source. 
Accretion is likely common  in post-asymptotic giant branch (AGB) binary interactions but how it operates 
and how its consequences depend on binary separation remain open questions. 
Here we use high resolution global 3D  hydrodynamic simulations of CEE with the adaptive mesh refinement (AMR) code AstroBEAR, 
to bracket the range of CEE companion accretion rates by comparing runs that remove mass and pressure via a subgrid accretion model with those that do not.  
The results show that if a pressure release valve is available, super-Eddington accretion may be common. 
Jets are a  plausible release valve in these environments, and they could also help unbind and shape the envelopes.
\end{abstract}
\begin{keywords}
binaries: close -- accretion, accretion discs -- stars: kinematics and dynamics -- hydrodynamics -- methods: numerical
\end{keywords}

\defcitealias{Ricker+Taam08}{RT08}
\defcitealias{Ricker+Taam12}{RT12}
\defcitealias{Ohlmann+16a}{ORPS16}
\defcitealias{Ohlmann+17}{ORPS17}
\defcitealias{Krumholz+04}{KMK04}
\section{Introduction}
\label{sec:introduction}
Close binary star interactions lie at the heart of many interesting and poorly-understood stellar astrophysical phenomena.  
Common envelope evolution (CEE), whereby a binary pair rapidly inspiral as the secondary enters the outer layers of the primary, 
represents such a  binary process that can lead to a variety of crucial phenomena in stellar evolution \citep{Paczynski76,Iben+Livio93,Ivanova+13,Demarco+Izzard17}.  

Binaries are, for example, likely needed to explain the ubiquity of bipolar planetary nebulae (PNe) and pre-planetary nebulae (PPNe) 
\citep{Soker94,Reyes-Ruiz+1999,Soker+2000,Soker+2001,Blackman+2001,Balick+Frank02,Nordhaus+Blackman06,Nordhaus+07,Witt+2009}.  
PNe collimated outflow momenta  are  much less kinematically demanding than those found in PPNe so far \citep{Bujarrabal+01,Blackman+2014,Sahai+2017}.
If PNe are the evolved states of PPNe, 
strongly (rather than weakly) interacting binaries and the associated modes of accretion may be essential to explain these high momenta
\citep{Blackman+2014}.

The binary central stars of several bipolar PNe are close enough to imply that they, and likely many more PNe, experienced a common envelope interaction phase. 
The ejection of the primary's envelope, which is expected to be a necessary consequence of CEE, will likely play a pivotal role in the formation of PNe. 
Winds from the exposed primary core (a proto white dwarf) will be shaped by their inertial interaction with toroidal ejected envelope 
and this may be a fundamental mechanism for producing PN bipolar morphologies \citep[][and references therein]{Jones+Boffin17}. 

For more massive binary stars, 
progenitors of gravitational wave (GW) generating mergers likely pass through a CEE  phase \citep{Kalogera+07,Ivanova+13,Belczynski+14}. 
The black hole (BH)-BH and neutron star (NS)-NS sources of recent gravitational wave detections 
via the Laser Interferometer Gravitational-Wave Observatory (LIGO)
(and other GW detectors) were most likely preceded by a CEE phase that set-up the conditions for mergers
\citep[e.g.][]{Abbott+16,Abbott+17a,Abbott+17b}.

Although first proposed more than four decades ago \citep{Paczynski76},  
much of the physics of CEE still remains uncertain. 
The process involves inherently 3D fluid dynamics
(and magnetic fields; \citealt{Nordhaus+Blackman06,Nordhaus+07})
but early analytic formalisms for CEE employ  simplified  parameterizations for energy or angular momentum exchange/loss 
\citep{Livio+Soker88,Dekool90,Iben+Livio93,Nelemans+00,Webbink08,Ivanova+13}.  
Numerical studies of CEE were, likewise, hampered by the need for both full 3D models and high resolution.  

Early simulations of CEE include \citet{Rasio+Livio96,Sandquist+98,Sandquist+00,Lombardi+06}.
Over the last decade, more numerical codes have been adapted to study CEE.  
These include both smoothed particle hydrodynamics (SPH) and grid-based (often adaptive or moving mesh) models. 
Beginning with \citet{Ricker+Taam08,Ricker+Taam12} and \citet{Passy+12a} the community now has an expanding array of tools to study CEE.  
The results are so far encouraging and puzzling.  
While the early inspiral phase has been recovered in a variety of studies 
\citep{Nandez+14,Ohlmann+16a,Staff+16,Kuruwita+16,Ivanova+Nandez16,Iaconi+17a,Iaconi+17b}
almost all models show the orbital decay flattening out at distances too large to account for observations. 
Likewise the ejection of the envelope has proven difficult to achieve as much of the mass set in motion by the inspiral fails to reach the escape velocity 
and hence would tend to fall back \citep{Ohlmann+16a,Kuruwita+16}. 
This behavior is seen in all models to date,  and a number of explanations have been proposed.  
Mechanisms which allow the stars to continue to draw closer 
may operate on longer timescales than the simulations (ie. thermal or stellar evolutionary timescales). 
Others have proposed that mechanisms {\it not included} in the initial studies can drive the envelope away and allow the binary orbit to continue to shrink.  
Such mechanisms include recombination \citep{Nandez+15,Ivanova+Nandez16} in the expanding/cooling envelope 
or radiation pressure on dust grains \citep{Glanz+Perets18}.  
The efficacy of such mechanisms remains strongly debated \citep[e.g.][]{Grichener+18}.

Accretion  in CEE is of interest both because it 
may have a role in the envelope ejection and also because it is a ubiquitous
engine for outflows \citep{Frank+02}.
As discussed above, such outflows are prevalent in PPNe and PNe
but  have also been considered as a means for driving some types of supernova \citep{Milosavljevic+12,Gilkis+16}. 
If such outflows occur during common envelopes (CEs) either via accretion onto the primary core \citep{Blackman+2001,Nordhaus+11} or onto the secondary, 
the evolution may be altered and perhaps drive more of the envelope to escape velocity.  
Some previous studies attempted to characterize accretion in CEE simulations as part of global AMR simulations \citep{Ricker+Taam08,Ricker+Taam12}, 
and in ``wind tunnel''  formulations \citep{Macleod+Ramirez-ruiz15b,Macleod+17}.
They found that accretion did occur with rates that were below that of Bondi-Hoyle-Littleton (BHL) flows 
\citep[][see \citealt{Edgar04} for a review]{Hoyle+Lyttleton39,Bondi+Hoyle44,Bondi52}, but still super-Eddington.
\citet{Murguia-berthier+17} also used a ``wind tunnel'' formulation to explore CEE accretion 
and found that for lower values of the polytropic index $\gamma$ accretion discs could form.

In this work we introduce and use a new tool to carry out CEE simulations, 
with a particular focus on accretion around the secondary.  
Using  our AMR MHD multi-physics code AstroBEAR \citep{Cunningham+09,Carroll-nellenback+13},%
\footnote{For a discussion of angular momentum conservation in AstroBEAR, we refer the reader to \citet{Blank+16}.}
we have developed modules for simulating CEE 
and here we describe our initial results following the inspiral of a red giant branch (RGB) star and a smaller companion.  
We present results from two high-resolution simulations, 
one of which uses a subgrid module for accretion onto the ``sink particle'' secondary star 
that removes mass and pressure \citep{Krumholz+04} (hereafter, \citetalias{Krumholz+04}), 
and the other without a subgrid model. 
Both cases  show general features of the inspiral but a dramatic difference in accretion rates between the two aforementioned cases
highlights how very different conclusions about CEE accretion can be reached depending on the presence or absence of an inner loss valve.

In Section \ref{sec:methods} we describe our method.  
In Section \ref{sec:results} we present results from the two aforementioned simulation cases, focusing on disk formation and accretion.
In Section \ref{sec:discussion} we discuss the differences that these two cases imply for the role of accretion in CEE, 
and what is required to sustain accretion onto the companion. 
In Section \ref{sec:convergence} we summarize some numerical challenges and we conclude in Section~\ref{sec:conclusions}.

\section{Methods}
\label{sec:methods}
\subsection{Setup}
\label{sec:setup}
We solve the equations of hydrodynamics for a binary system consisting of a red giant (RG) and an unresolved stellar companion represented by a (gravitation only) sink particle with mass equal to half of the RG mass. 
We adopt an ideal gas equation of state with adiabatic index $\gamma=5/3$.
Gravitational interactions between particles, and between particles and gas, as well as self-gravity of the gas, are calculated self-consistently.
Although our numerical setup and chosen physical parameters follow closely those of \citet{Ohlmann+16a,Ohlmann+17} 
(hereafter \citetalias{Ohlmann+16a} and \citetalias{Ohlmann+17}, respectively),
the numerical methods are very different (e.g. our AMR vs. their moving mesh).
In particular, our RG model and setup is very similar to theirs 
with a few minor differences discussed below.
The similarity was deliberate because this RG setup resulted in a star very close to hydrostatic equilibrium
and enables a consistency check between our independently obtained results and theirs.

The reader is referred to \citetalias{Ohlmann+17} for details, but we summarize
the procedure and notable differences between the two approaches below.
We first evolve a star with a zero-age main sequence (MS) mass of $2M_\odot$ 
using the 1D stellar evolution code MESA (version~8845) \citep{Paxton+11,Paxton+13,Paxton+15}, setting the metallicity $Z$ to $0.02$, 
and select the snapshot that most closely coincides with the RG of \citetalias{Ohlmann+16a,Ohlmann+17} on the Hertzsprung-Russell diagram.
We call this star  the ``primary'' and its companion  the ``secondary.''

Numerically resolving the pressure scale-height in the core is unfeasible.
We therefore truncate the RG at a radius $r=r\cut=2.41R_\odot$, 
and replace the core by the combination of a gravitation-only sink particle and a surrounding density profile  which smoothly matches the density at $r\cut$.
The modified profile is obtained by numerically solving a modified Lane-Emden equation,
with polytropic index $n=3$, taking into account the gravitation of the sink particle
and  boundary conditions for $\rho$ and $d\rho/dr$. 
This particle is  the ``primary particle'' and the remainder of the RG is the ``primary envelope.'' 
Their masses are $m_1$ (primary particle) and $m\oneenv$ (primary envelope)
where the total primary mass $M_1=m_1 + m\oneenv$.
Unlike in \citetalias{Ohlmann+16a,Ohlmann+17}, where $m_1$ is set equal to the interior mass $m(r\cut)$ of the MESA profile,
we iterate over $m_1$, solving the equation at each iteration until $m_1+m\oneenv(r\cut)=m(r\cut)$, 
where $m(r)$ is the interior mass and $m\oneenv(r\cut)$ is the interior gas mass of the modified profile.
This prevents the mass of the modified RG from exceeding that of the original MESA model
and, more importantly,  maintains a higher degree of hydrostatic equilibrium in the RG than would have otherwise obtained.
The initial mass and radius of the primary are $M_1= 1.956M_\odot$ and $R_1= 48.1R_\odot$, respectively,
with $m_1=0.369M_\odot$.

Simulations are carried out in the inertial centre of mass frame, 
but with the centre of the mesh coinciding with the initial position of the primary particle.
We choose extrapolating hydrostatic boundary conditions and adopt a multipole expansion method for solving the Poisson equation.
The ambient medium is chosen to have a constant density and pressure of $6.67\times10^{-9}\gcmcmcm$ and $1.01\times10^5\dynecmcm$, values 
similar to those at the surface of the RG. 
An ambient pressure (seven orders of magnitude smaller than the central pressure of the modified envelope) 
is added everywhere in the domain to obtain a smooth transition between the stellar surface and its surroundings and to ensure that the pressure scale-height is adequately resolved at the stellar surface.
Using a lower ambient density results in larger ambient sound speeds, smaller time-steps, and hence reduced computation speeds.
In lower resolution tests, we found that reducing the ambient density to $10^{-10}\gcmcmcm$
makes an insignificant difference to our results (see Appendix~\ref{sec:test_runs}).
We also experimented with  a hydrostatic atmosphere instead of a uniform ambient medium,
but this was numerically unstable at the corners of the mesh.

We place a second sink particle with mass equal to half that of the RG, or $m_2=0.978M_\odot$, 
at a distance $a\f=49.0R_\odot$ from the primary  particle, just outside of the RG, at $t=0$. 
This secondary particle represents either a MS star or a white dwarf (WD).
For both particles, we used a spline function \citep{Springel10}  with softening length $r\soft$ set  equal to $r\cut$.
The particles and the RG envelope are initialized in a circular Keplerian orbit.
We  initialized the RG with zero spin relative to the centre of mass frame.
This differs from \citetalias{Ricker+Taam08,Ricker+Taam12} and \citet{Ohlmann+16a}, for example,
where the envelope is initialized with a solid body rotation of $0.95$ times the initial orbital angular velocity.
A more realistic estimate might be $\sim0.3$ \citep{Macleod+18}.%
\footnote{\citet{Macleod+18} simulate the phase starting with Roche lobe overflow and ending with plunge-in,
with initial separation equal to the Roche limit estimated analytically from \citet{Eggleton83}.
They initialize the primary to spin rigidly in corotation with the orbit.
When the inter-particle separation equals the initial radius of the primary,
the spin of the primary almost equals its initial spin at the Roche limit separation.
If we adopt this for our binary system and apply the analytic estimate of the Roche limit used by \citet{Macleod+18},
then we obtain a spin at $t=0$ of $30$ per cent of the instantaneous orbital angular velocity.
}

Below we compare two runs called  Model~A and Model~B. 
The essential difference is that a subgrid accretion model is implemented only for Model~B
which removes mass and pressure.
The setups for these runs are otherwise only slightly different:
Model~A uses a box with side length $L=1150R_\odot$, while for Model~B $L=575R_\odot$.
For Model~B, we apply the velocity damping algorithm of \citetalias{Ohlmann+17} until $5t\dynam$, with $t\dynam$ set to $3.5\da$,
but for Model~A we do not apply any velocity damping.
The (pre-$t=0$) relaxation run with damping used for Model~B is carried out with the same box size as for Model~B, 
and with resolution equal to the initial resolution of Model~B.
This produces minor differences in the initial conditions between the two runs.
But the close correspondence of the orbits 
up to when the accretion rate in Model~B becomes significant at $t\sim14\da$ (see Section~\ref{sec:accretion}),
the striking similarity in density snapshots of Figures~\ref{fig:rho_143} and \ref{fig:rho_132} at $t=0$ and $t=10\da$, 
and the rather sudden emergence of differences in the orbits/morphologies shortly after $t=14\da$,
show that any differences in the results caused by the small differences in initial conditions 
(and box size and refinement algorithm; see below) 
are negligible in comparison with the differences caused by the presence/absence of the accretion subgrid model.

The highest spatial resolution of $0.140R_\odot$ and base resolution of the ambient volume of $2.25R_\odot$ are the same for Models~A and B,
and there is a buffer zone in between to allow the resolution to transition gradually.
The region within, of maximum refinement by the code, is slightly different in extent and shape
for Models~A and B, as are the extents of the buffer zones.\footnote{For Model~A, 
this region is spherical and centred on the primary particle until $t=16.7\da$, after which it is centred on the secondary.
For Model~B, there are two such overlapping regions, one spherical centred on the primary particle, 
and the other cylindrical with axis orthogonal to the orbital plane and centred on the secondary.}
For both runs however, the moving region of maximum refinement contains the particles and
 a portion of the surrounding gas  at all times, so that the resolution is both uniform and  high in the region of interest.
In addition, the softening length for the sink particles is reduced to half of its initial value 
about halfway through the simulation for Model~A, and simultaneously the smallest resolution cell is halved to $0.070R_\odot$, but not for Model~B.
This ensures that the softening length never exceeds a fraction of $1/5$ of the inter-particle separation \citepalias[cf.][]{Ohlmann+16a}.
Limited computational resources prohibit us from redoing one of the runs to make these parameter values match more precisely,
but we are confident that these differences are inconsequential  compared to the presence
or absence of the subgrid accretion model and do not affect our conclusions.
Finally, Model~A is run up to $40\da$, while Model~B is run up to $69\da$ but we choose to present results for the first $40\da$ only.

\subsection{Modelling the accretion}
\label{sec:accretion_setup}
Model~A does not employ a subgrid accretion model, and thus resembles closely the setup of \citetalias{Ohlmann+16a},
and, to a lesser extent those of the other global CE simulations from the literature which also do not have subgrid accretion.
Model~B employs the accretion model of \citetalias{Krumholz+04} for the secondary,
but not for the primary particle, because our goal is to explore accretion onto the secondary.%
\footnote{We shall see in Section~\ref{sec:accretion} 
that while the flow around the secondary has certain properties expected for an accretion flow even in Model~A (no subgrid accretion), 
the same cannot be said about the flow around the primary particle.}
This prescription is based on the BHL formalism \citep[][see \citealt{Edgar04} for a review]{Hoyle+Lyttleton39,Bondi+Hoyle44,Bondi52}.

\citet{Ricker+Taam08,Ricker+Taam12,Macleod+Ramirez-ruiz15a,Macleod+17}
found that BHL accretion overestimates  the accretion rate in CE evolution, 
which is not unexpected given that the conditions of the problem violate the assumptions of the BHL formalism \citep{Edgar04}.
Our focus is not on this point, but rather on comparing the accretion and evidence for
disc formation from a simulation that
allows accretion onto the secondary (Model B) using the \citetalias{Krumholz+04} model with one that does not (Model A).
Although the \citetalias{Krumholz+04} prescription was not designed for the present context (as we discuss further later) 
it is  well-tested numerically \citep{Li+14}, and is currently the best tool we have for this purpose.

Accretion is permitted to take place within a zone of four grid cells from the secondary.
\citetalias{Krumholz+04} suggests that the \textit{Plummer} softening radius should be smaller or equal to the accretion radius,
to avoid artificially reducing the accretion rate due to the reduced gravitational acceleration inside the softening sphere.
The spline potential employed is roughly equivalent to a Plummer potential with a Plummer softening radius that is $2.8$ times smaller
than the spline softening radius of $\approx17$ grid cells (this factor gives equal values of the potential at the origin).
Thus, the accretion radius (4 cells) is  slightly smaller than the Plummer-equivalent softening radius ($\approx6$ cells),
and likely slightly reduces the accretion rate compared to when the two radii are equal (see also Appendix~\ref{sec:test_runs}).
This makes our subgrid model a slightly ``milder'' version of \citetalias{Krumholz+04}.

\section{Results}
\label{sec:results}

\begin{figure*}
  \includegraphics[height=50mm,clip=true,trim= 50 40 250 160]{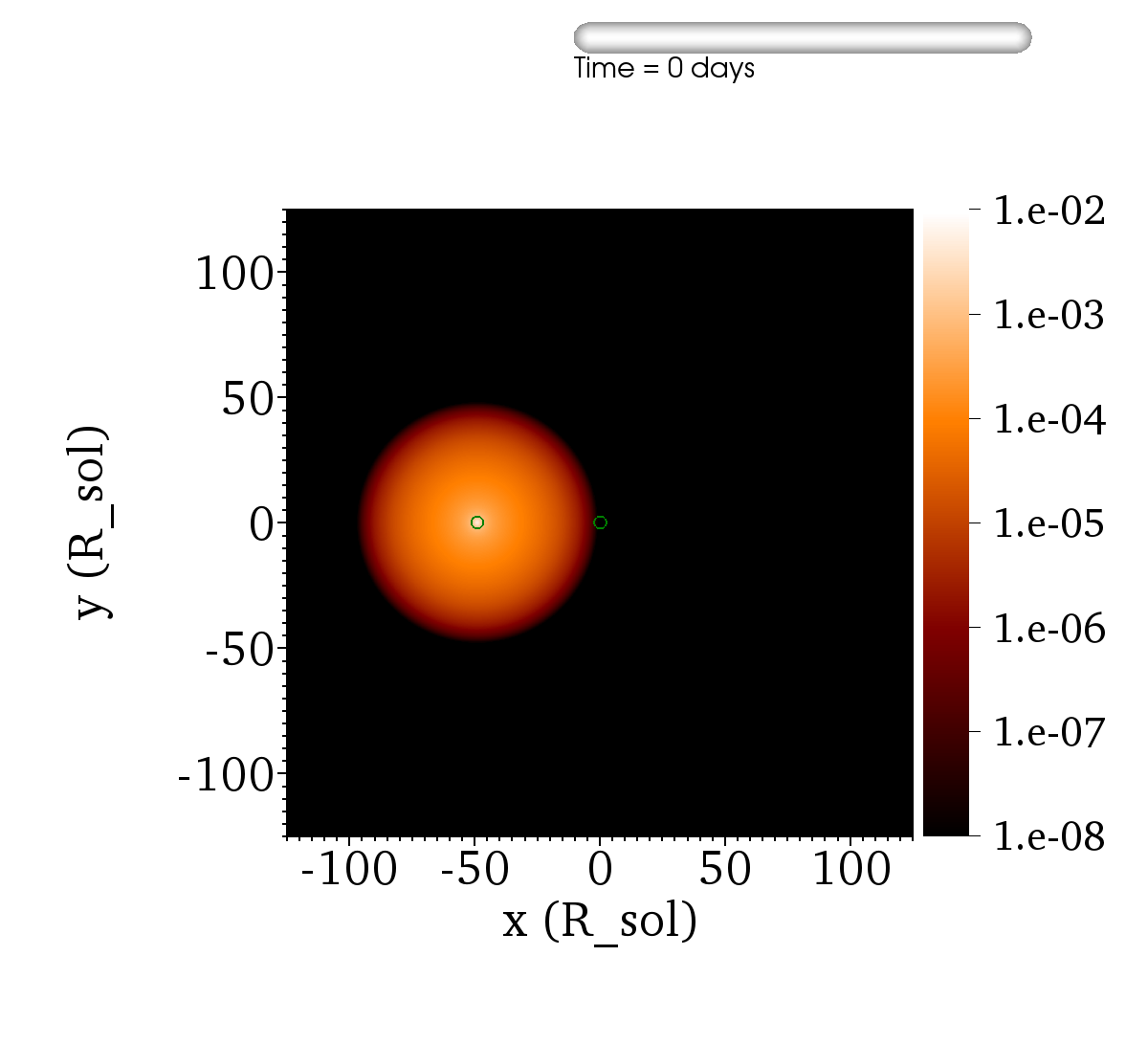}
  \includegraphics[height=50mm,clip=true,trim=290 40 250 160]{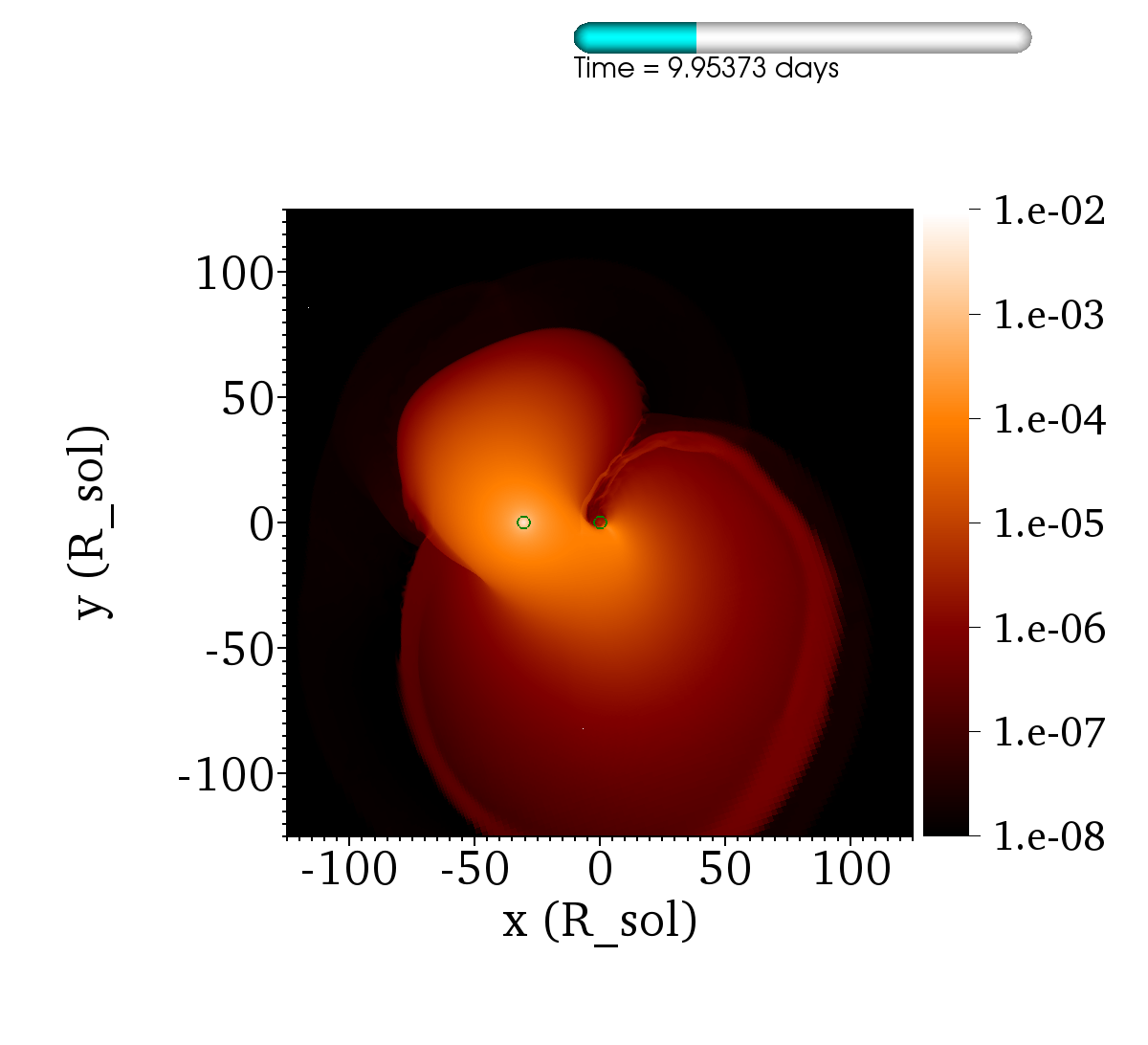}
  \includegraphics[height=50mm,clip=true,trim=290 40 250 160]{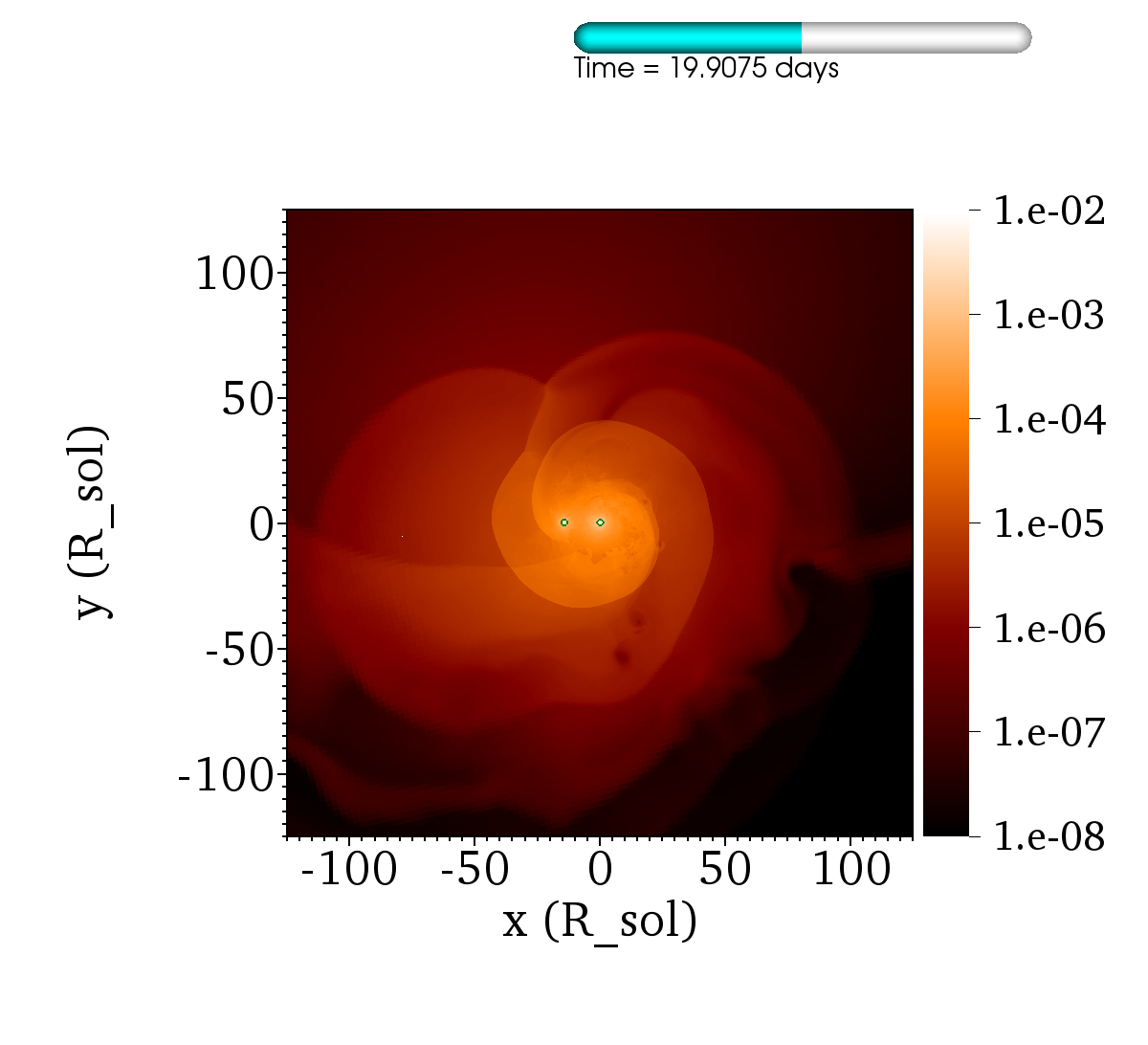}
  \includegraphics[height=50mm,clip=true,trim=290 40  40 160]{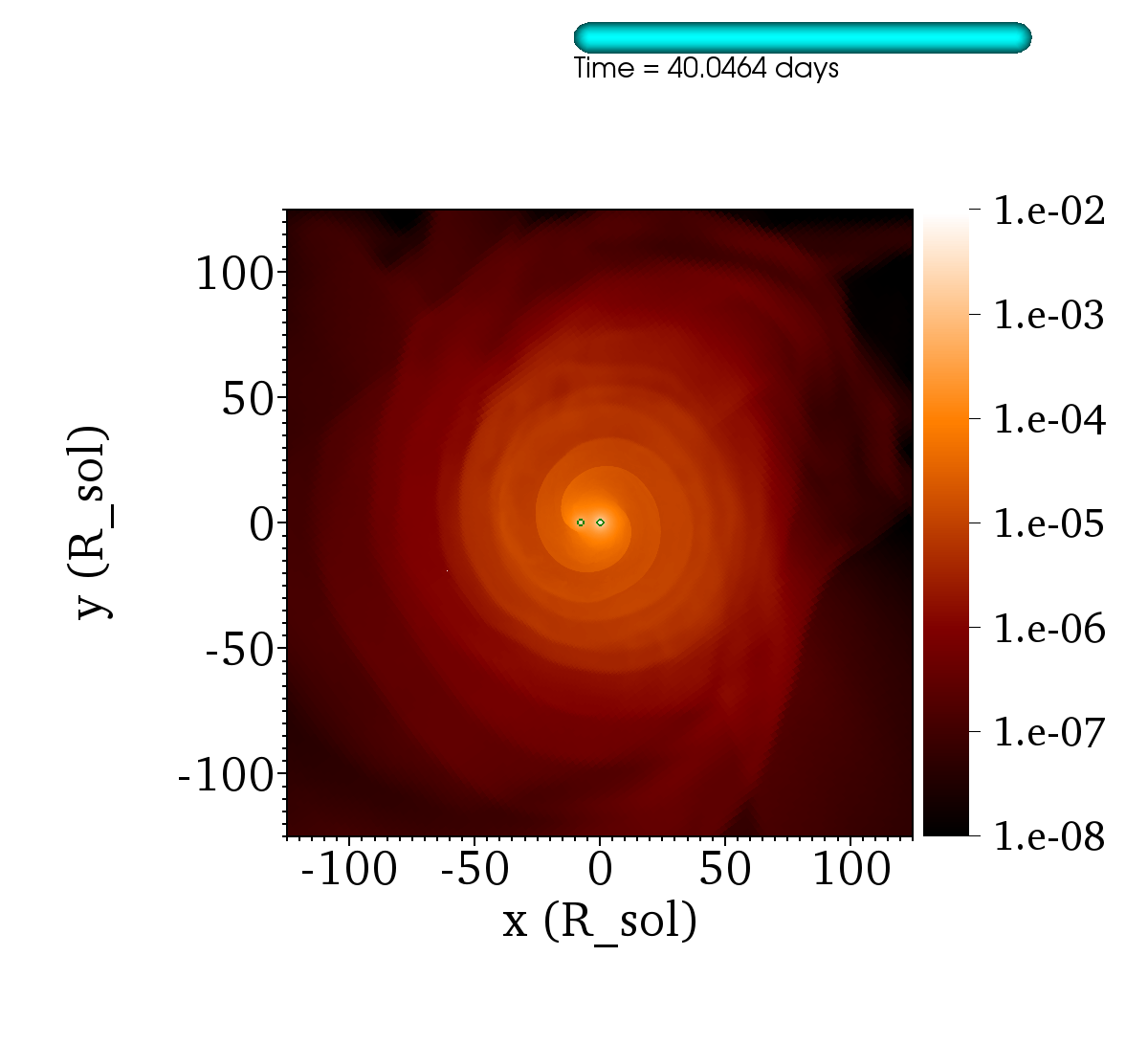}
  \caption{Density, in $\!\gcmcmcm$ in a slice through the secondary and parallel to the $xy$ (orbital) plane, 
           for Model~A (no subgrid accretion model).
           The secondary is positioned at the centre with the frame rotated so that the primary particle is always situated to its left
           (the plotted frame of reference is rotating with the instantaneous angular velocity of the particles' orbit).
           Both particles are denoted with a green circle with radius equal to the spline softening length.
           Snapshots from left to right are at $t=0$, $10$, $20$ and $40\da$. 
           \label{fig:rho_143}
          }            
\end{figure*}

\begin{figure*}
  \includegraphics[height=50mm,clip=true,trim= 50 40 250 160]{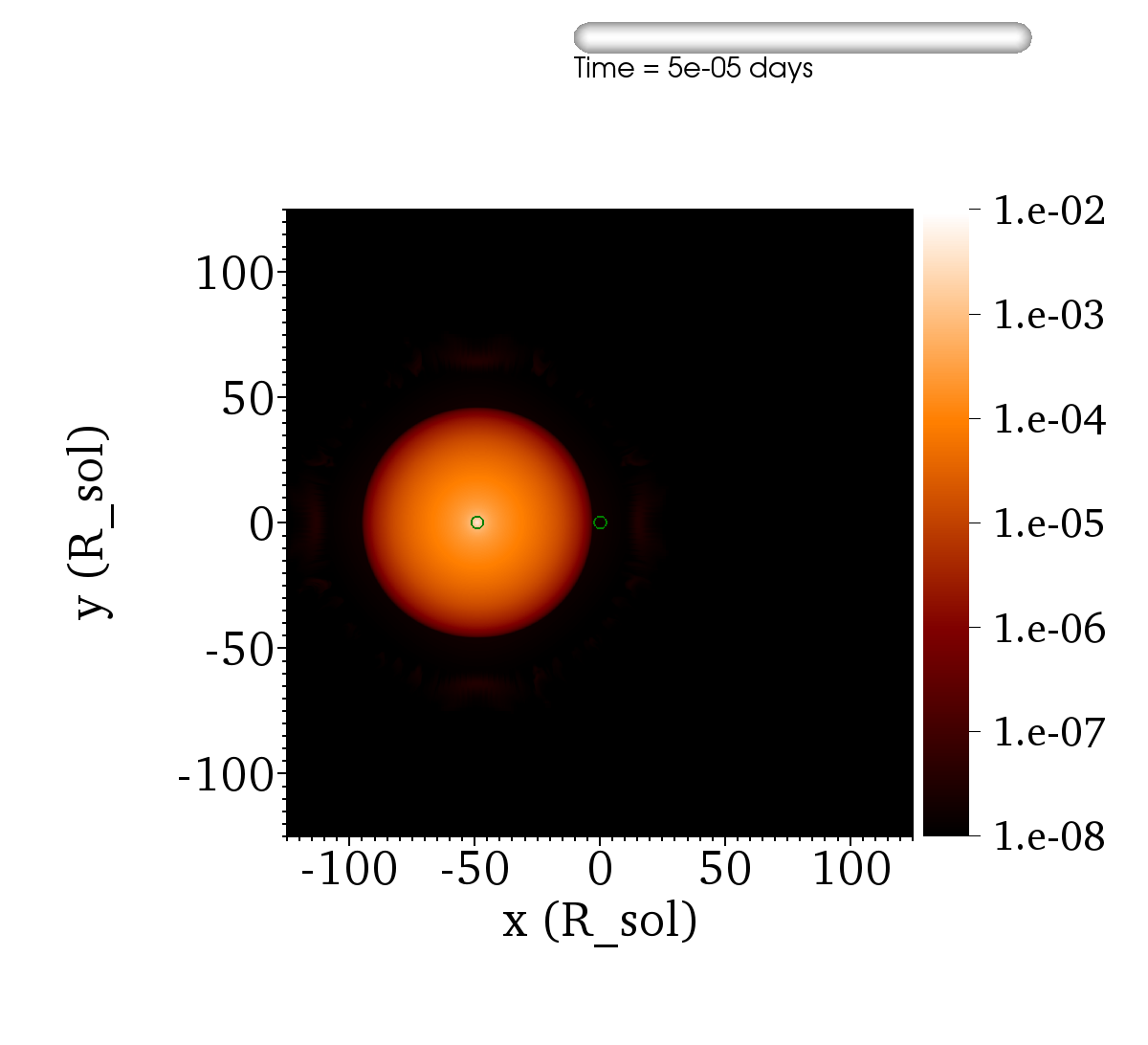}
  \includegraphics[height=50mm,clip=true,trim=290 40 250 160]{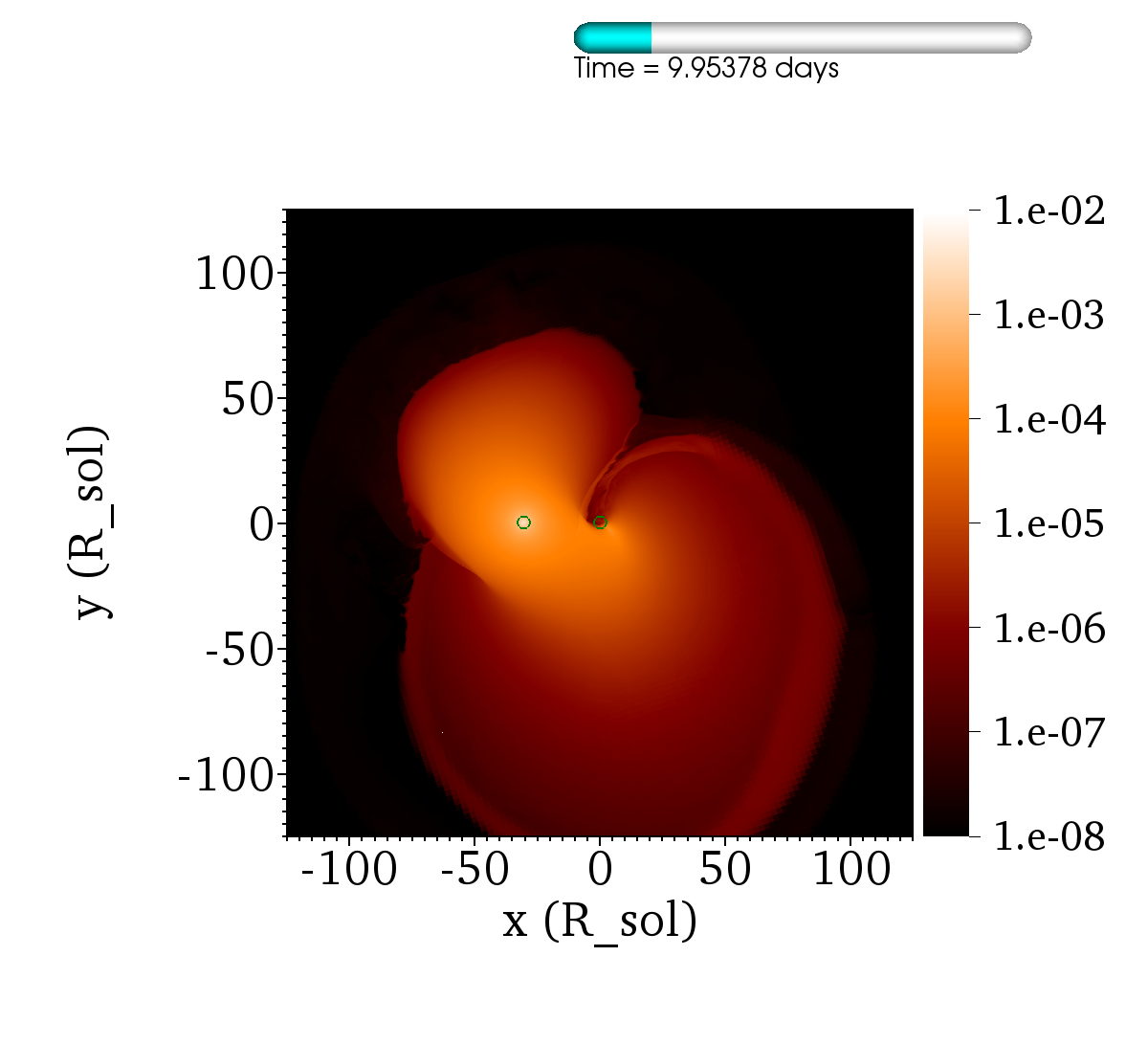}
  \includegraphics[height=50mm,clip=true,trim=290 40 250 160]{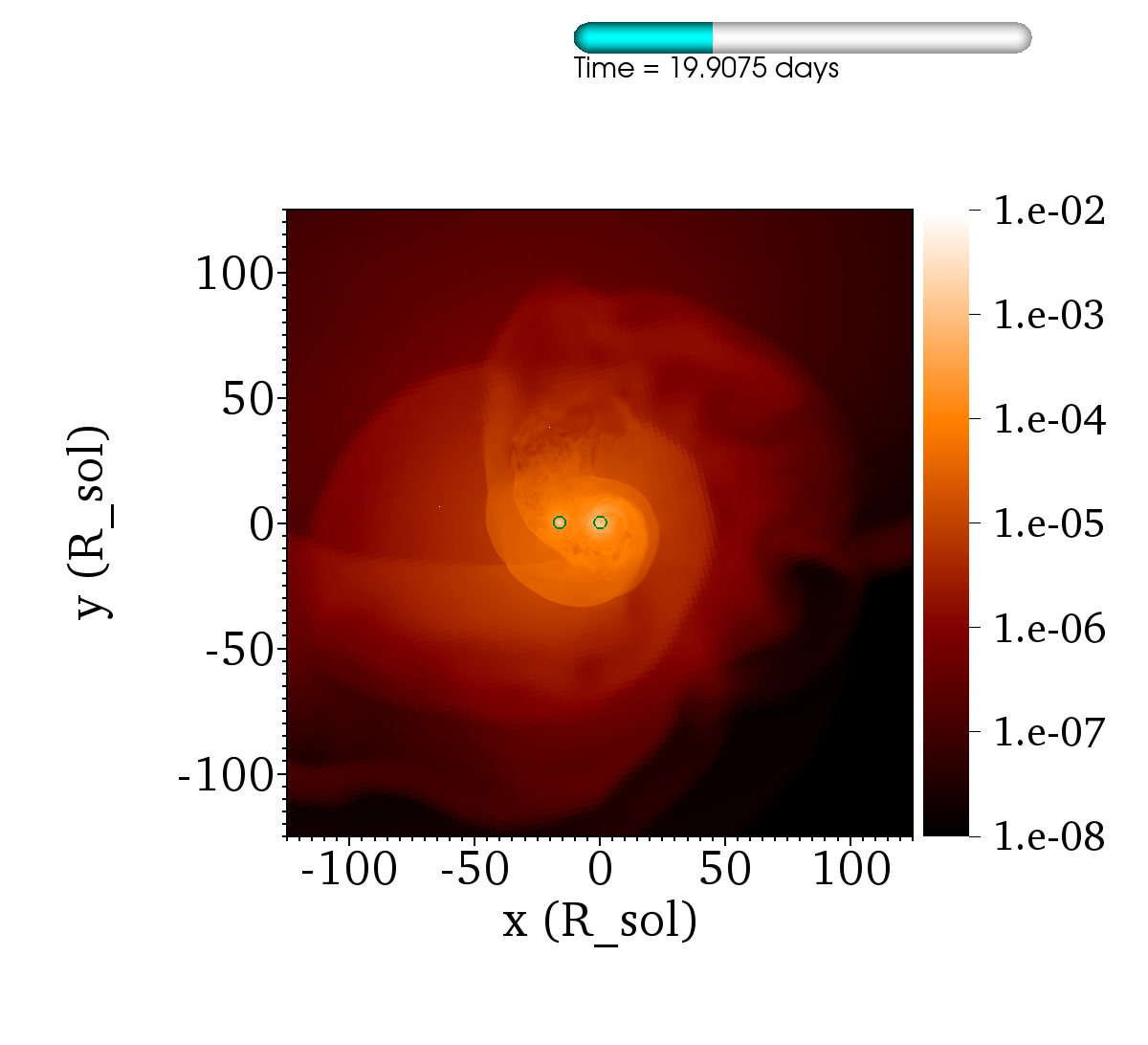}
  \includegraphics[height=50mm,clip=true,trim=290 40  40 160]{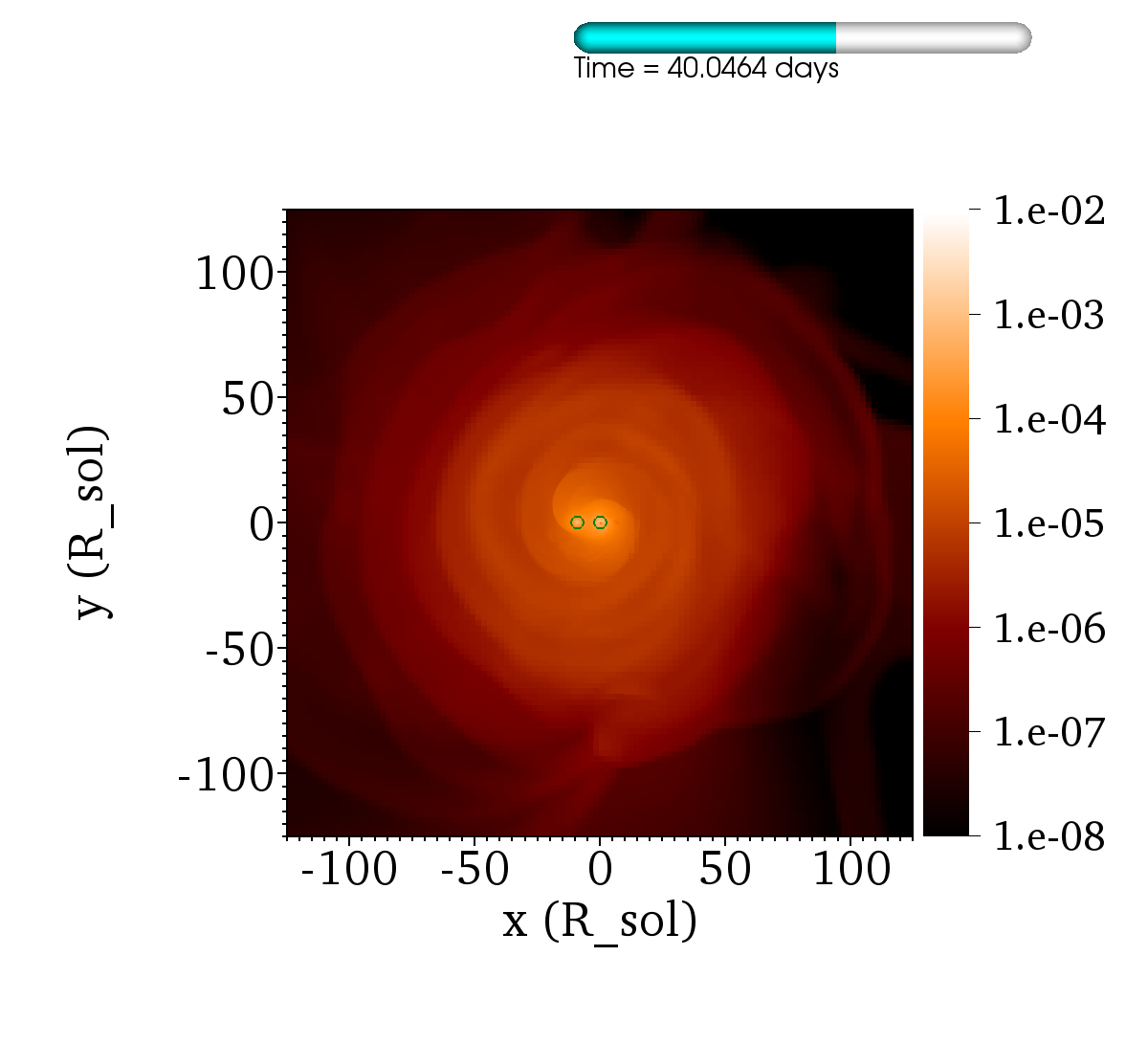}
  \caption{As Fig.~\ref{fig:rho_143} but now for Model~B, with the \citetalias{Krumholz+04} subgrid accretion model turned on for the secondary.
           \label{fig:rho_132}
          }            
\end{figure*}

\begin{figure*}
  \includegraphics[width=\textwidth,clip=true,trim= 0 0 0 0]{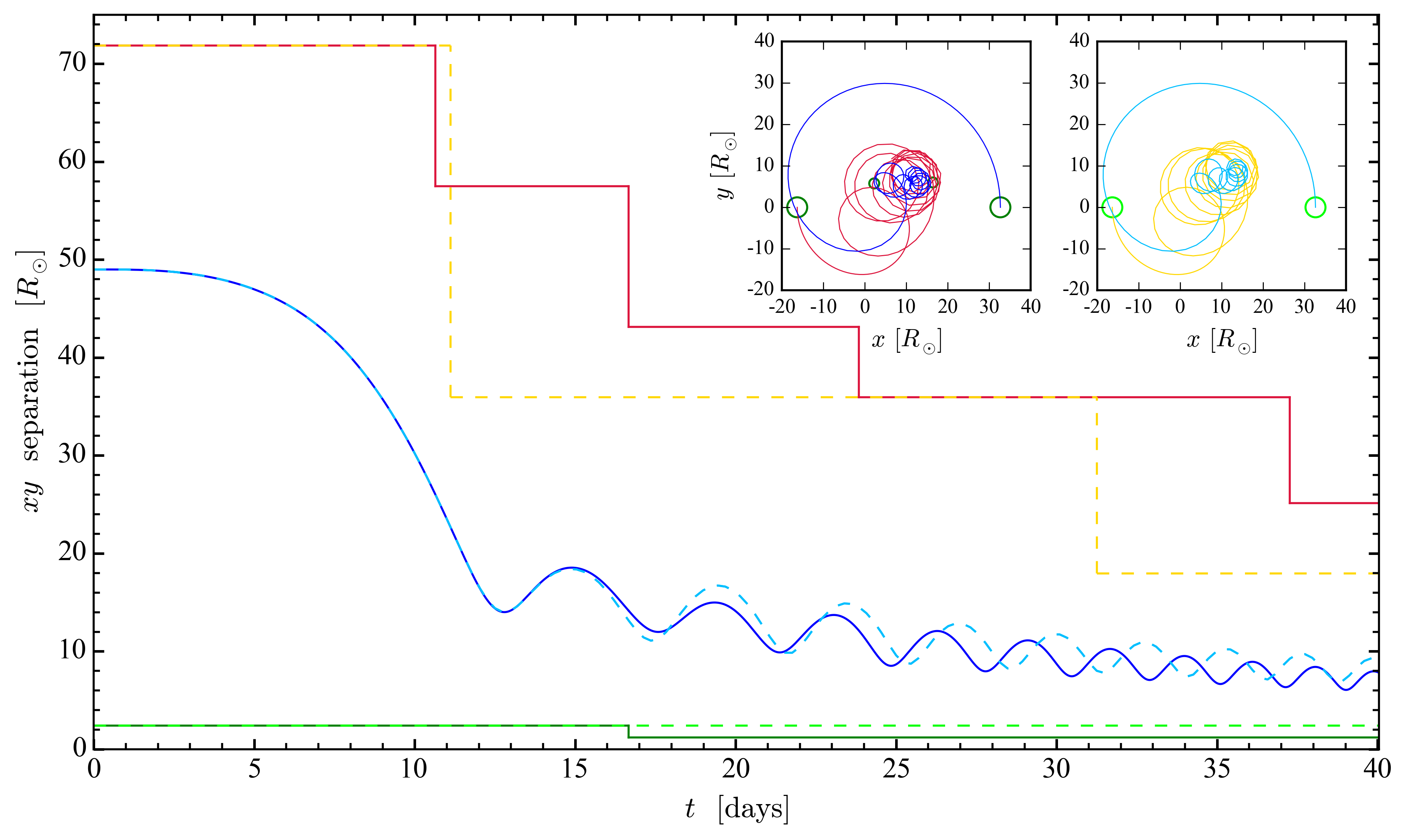}\\
  \caption{Inter-particle separation in the orbital plane ($z=0$) 
           for Model~A, without subgrid accretion (solid blue), and Model~B, with subgrid accretion (dashed light blue).
           Also shown are jagged lines denoting the radius of the spherical region of highest mesh refinement 
           (solid red for Model~A and dashed orange for Model~B),
           and the spline softening radius (solid green for Model~A and dashed light green for Model~B). 
           \textit{Inset}: Orbit of the sink-particles, with Model~A depicted on the left and Model~B on the right.
           The centre of mass is located at the origin in each panel.
           The primary particle is shown in red/orange while the secondary is shown in blue/light blue.
           Green/light green circles with radius equal to the spline softening length are shown at $t=0$
           and, for Model~A, also at $t=16.7\da$, when the softening length is halved.
           (The sampling rate used to draw the orbits is about one frame per $0.23\da$,
           resulting in a slightly ``choppy'' appearance at late times that is not related to the time sampling in the simulation.)
           \label{fig:orbit}
          }            
\end{figure*}

\subsection{Comparison of Morphological Properties and Inspiral}
\label{sec:runs}
 Figures~\ref{fig:rho_143} and \ref{fig:rho_132} show snapshots of slices of gas density in the orbital plane at $t=0$, $10$, $20$ and $40\da$,
for Models~A and B, respectively, with axes in units of $R_\odot$, and density in units of $\gcmcmcm$. 
In these figures, and others to follow, the secondary is located at the centre
and the primary particle is to its left, with the spline softening sphere depicted as a green circle around each particle.
These snapshots can be thought of as frames from a movie taken in a reference frame rotating with the instantaneous angular velocity of the particles.
The global evolution is very similar between the two runs, 
and closely agrees with the results of \citetalias{Ohlmann+16a}.
The spiral shock morphology that develops is also consistent with the results of other global CE simulations.

The distance $a$ between the two particles in the orbital plane as a function of time is illustrated in Figure~\ref{fig:orbit},
solid blue for Model~A and dashed light blue for Model~B.
Jagged solid red and dashed orange lines are plotted to show the radius of the sphere within which the resolution is at the highest refinement level,
while solid green and dashed light green show the spline softening radius, for Models~A and B respectively.
The initial reduction in separation for the first $\sim12\da$ is known as the plunge-in phase,
and each subsequent oscillation corresponds to a full orbit of $2\pi$ radians.

The curves for Models~A and B are almost identical up to $t\approx 15\da$, at which point they begin to diverge slightly.
This time does not correspond to any change in refinement radius or softening length,
but approximately to the time of peak in the accretion rate for Model~B, as will be discussed in Section~\ref{sec:accretion}.
The initial differences after $t=14\da$ are thus convincingly
caused by the difference in accretion prescriptions between the two runs.
In Model~A, 10 orbits are completed by $t=40\da$, while in Model~B, 9 orbits are completed, 
so the mean orbital frequency is higher in Model~A between $t=15\da$ and $t=40\da$ than for Model~B.
This is consistent with the mean inter-particle separation being slightly lower for Model~A than for Model~B during the same time interval. From $t\sim15$--$17\da$ however, Model~B shows a smaller separation and mean orbital period than Model~A and vice versa after $t\sim17\da$.
This suggests that the reduction in softening length at $t=16.7\da$ causes the orbital period to decrease in Model~A,
compared to what it would have been had the softening length remained the same, 
whereas the subgrid accretion causes a reduced orbital period  in Model~B between  $t\sim15$--$17\da$, 
from what it would have been had subgrid accretion been turned off.
Therefore, both subgrid accretion and reduction of the softening length tend to reduce the orbital period and mean separation.
We elaborate on this  in Sections~\ref{sec:accretion} and \ref{sec:discussion}.

Both curves of separation vs. time resemble that of \citetalias{Ohlmann+16a}.
However, in that work the particles complete only 7 orbits by $t=40\da$.
Moreover, their first minimum is lower than the second, which is not the case in our runs, 
where the minima and maxima decrease monotonically with time.
Furthermore, the eccentricity of the orbit, which is related to the amplitude of the separation curve,
is larger in \citetalias{Ohlmann+16a}.
The main cause for these differences is probably 
that in \citetalias{Ohlmann+16a} the RG is initialized with a solid body rotation of 95\% corotation,
whereas in our case the initial angular rotation speed is zero,
but it would be interesting to explore the effects of initial spin in a future study.

We now turn to the insets of Figure~\ref{fig:orbit}, where the orbits are plotted for Model~A on the left and for Model~B on the right.
The orbit of the primary particle is shaded in red/orange and the secondary in blue/light blue.
The spline softening spheres are indicated with green circles for $t=0$ and, for Model~A, also for $t=16.7\da$, 
when the softening length is halved.
Orbits resemble qualitatively the orbit obtained by \citetalias{Ohlmann+16a}.

\begin{figure*}
  \includegraphics[height=75mm,clip=true,trim= 50 120 220 170]{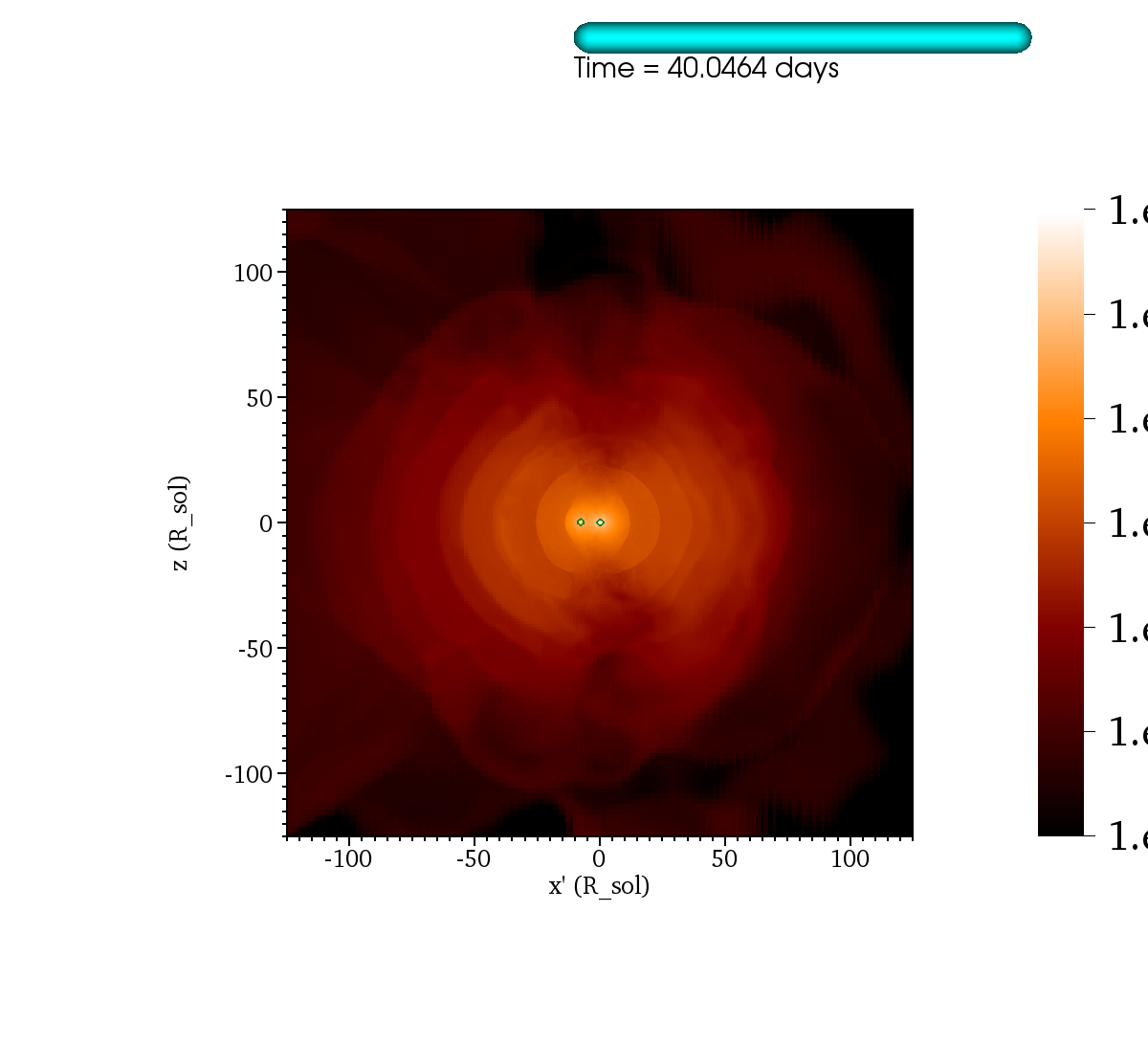}
  \includegraphics[height=75mm,clip=true,trim=220 120  40 170]{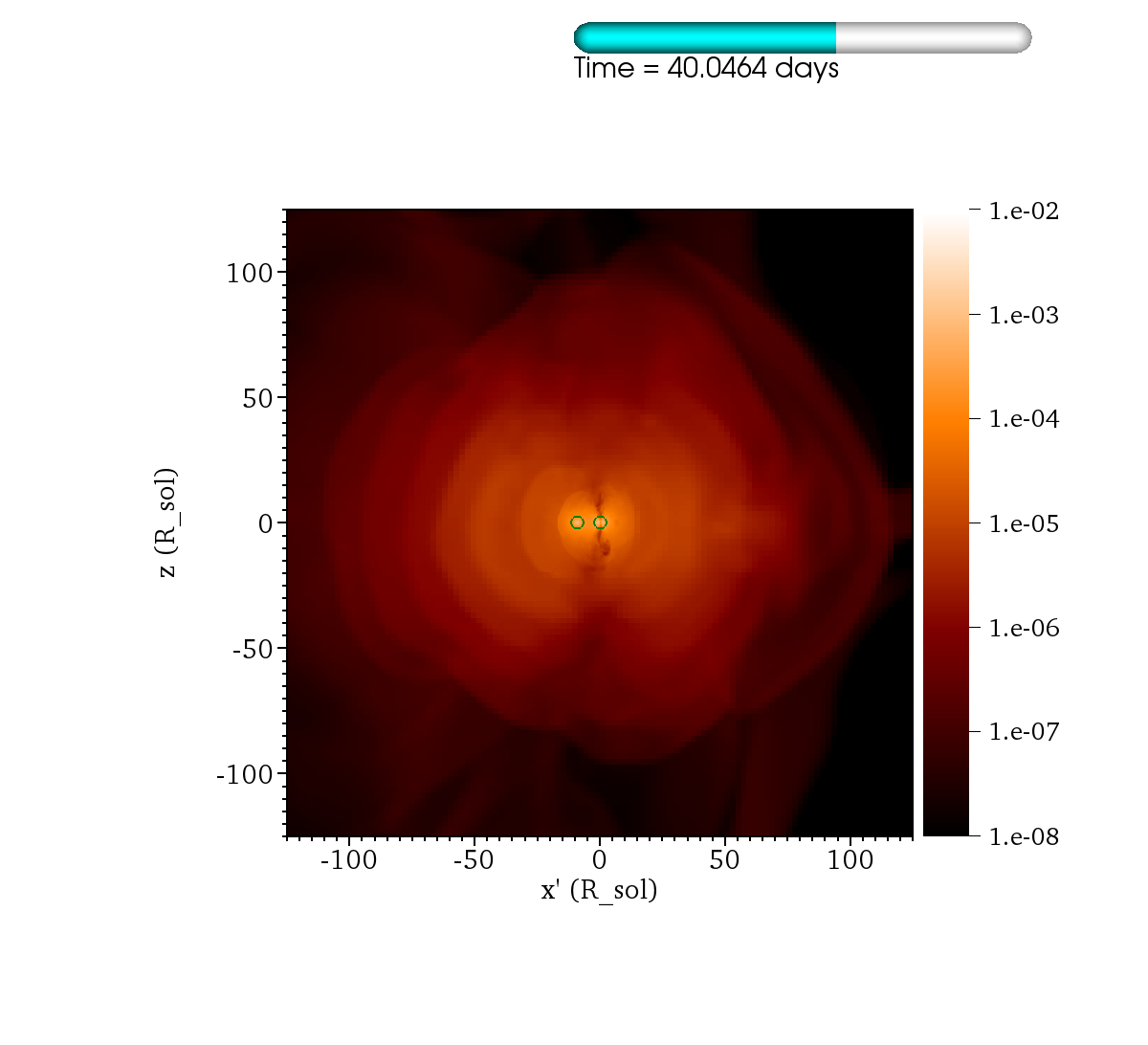}\\
  \includegraphics[height=75mm,clip=true,trim= 50 120 220 170]{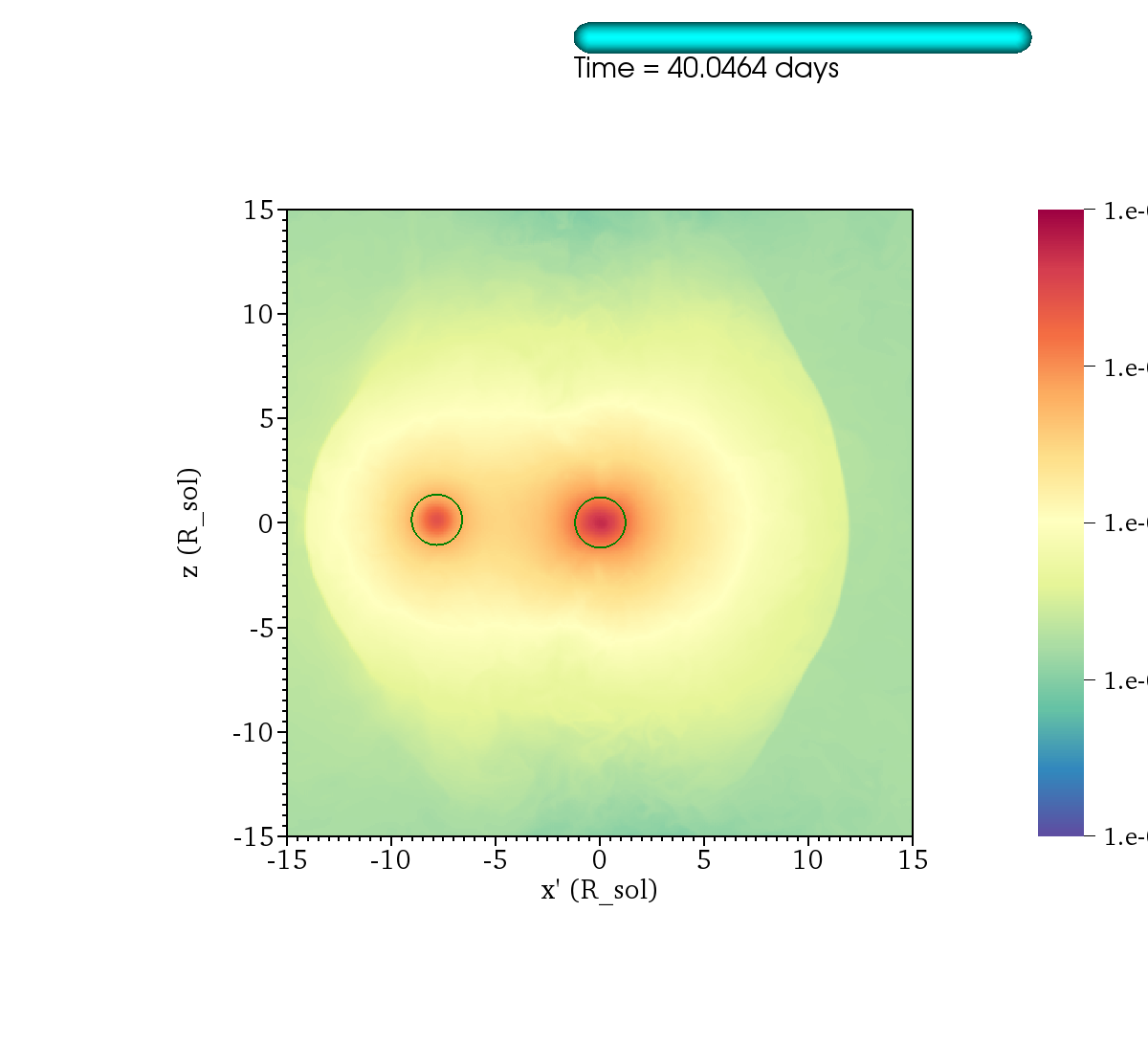}
  \includegraphics[height=75mm,clip=true,trim=220 120  40 170]{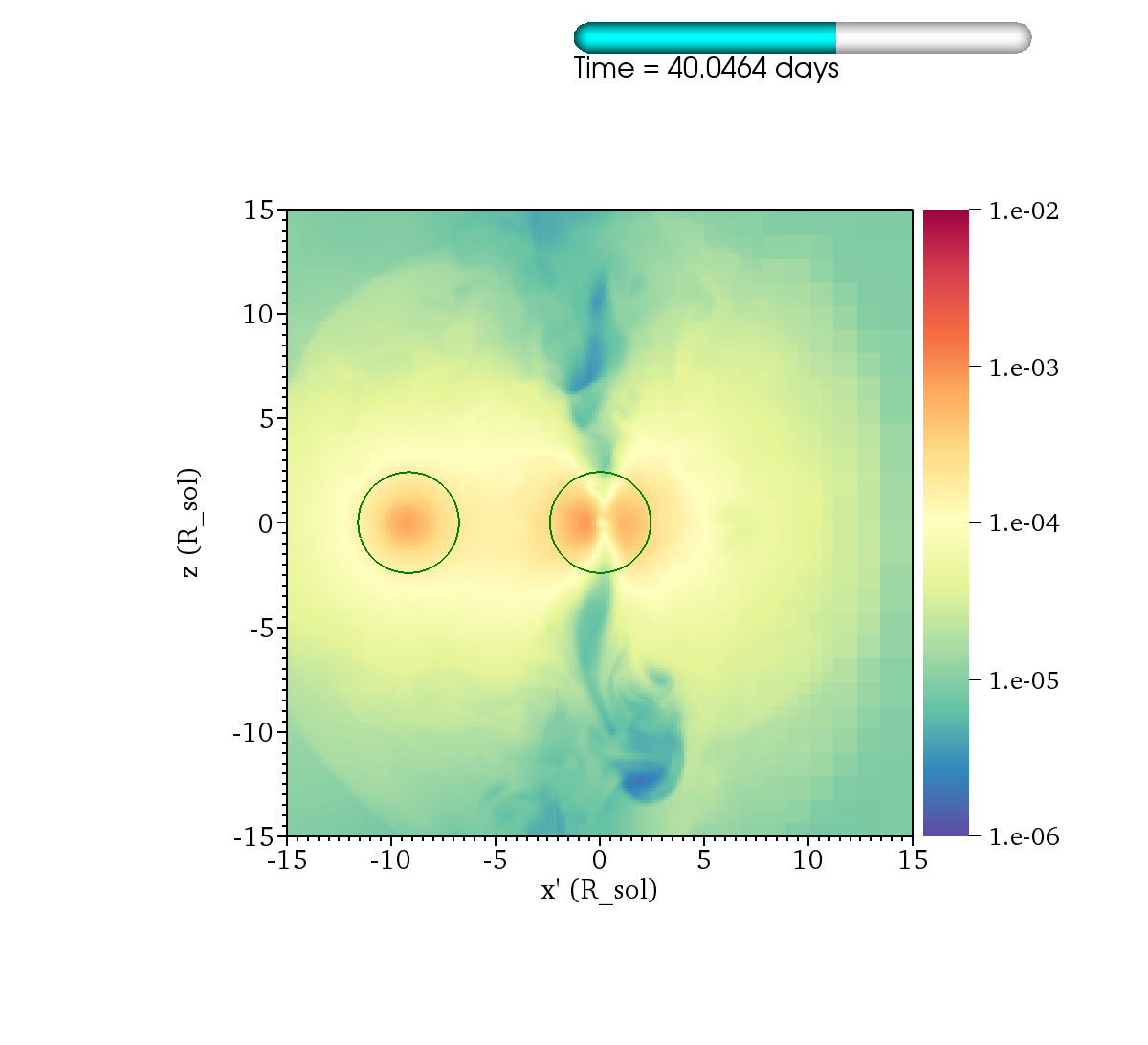}
  \caption{Gas density in $\gcmcmcm$ viewed in a slice through both particles that is perpendicular to the orbital plane at $t=40\da$.
           Model~A (no subgrid accretion) is shown in the left-hand column 
           and Model~B (\citetalias{Krumholz+04} subgrid accretion) is shown in the right-hand column,
           while the top and bottom row show two different levels of zoom (with different color schemes to aid viewing). 
           The secondary is situated in the centre of each panel and the frame is rotated so that the primary particle is located to its left.
           Spline softening spheres are identified by green circles.
           The $x'$-axis is defined by the line in the orbital plane that passes through both sink particles.
           \label{fig:rho_edgeon}
          }
\end{figure*}

Next we show, in Figure~\ref{fig:rho_edgeon}, slices of gas density $\rho$ at $t=40\da$ that pass through both particles 
and which cut through the orbital plane orthogonally, so that the view is edge-on with respect to the particles' orbit.
The left-hand column shows results for Model~A,  the right-hand column shows results for Model~B,
and the top and bottom rows present different levels of zoom (using different colour schemes for presentational convenience).
The layered shock morphology is qualitatively very similar to that seen in other CE simulations \citep[e.g.][]{Iaconi+17b}.

Models~A and B also show quite similar morphology but with one conspicuous difference.
A torus-shaped structure is present around the secondary in Model~B, which employs subgrid accretion. 
A much less pronounced  similarly shaped structure around the secondary is only marginally visible in Model~A (lower left panel;  
though inconspicuous, this structure is confirmed by its presence in other snapshots, not shown).
The toroidal structure in Model~B is suggestive  of a thick accretion disc and is accompanied by a low-density elongated bi-polar structure, 
seen in blue in the bottom-right panel of Figure~\ref{fig:rho_edgeon}.
Accretion is conspicuously the cause for the presence of this striking morphology in Model~B.
Below we examine the properties of the flow around the companion in more detail for both runs.

\begin{figure*}
  \includegraphics[width=\textwidth,clip=true,trim= 1.5 41.5 0.5  0]{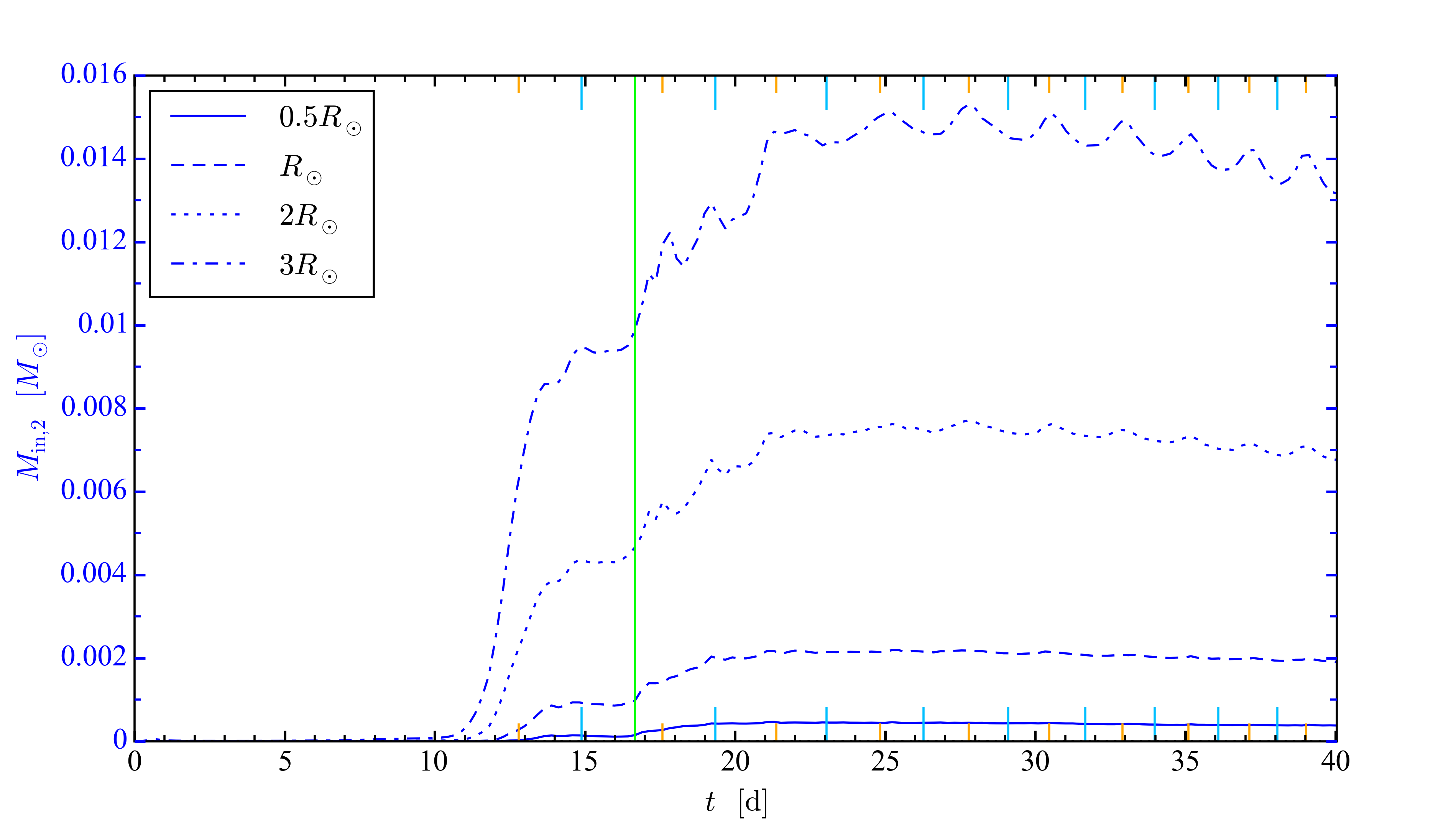} 
  \includegraphics[width=\textwidth,clip=true,trim= 0    0   0   15]{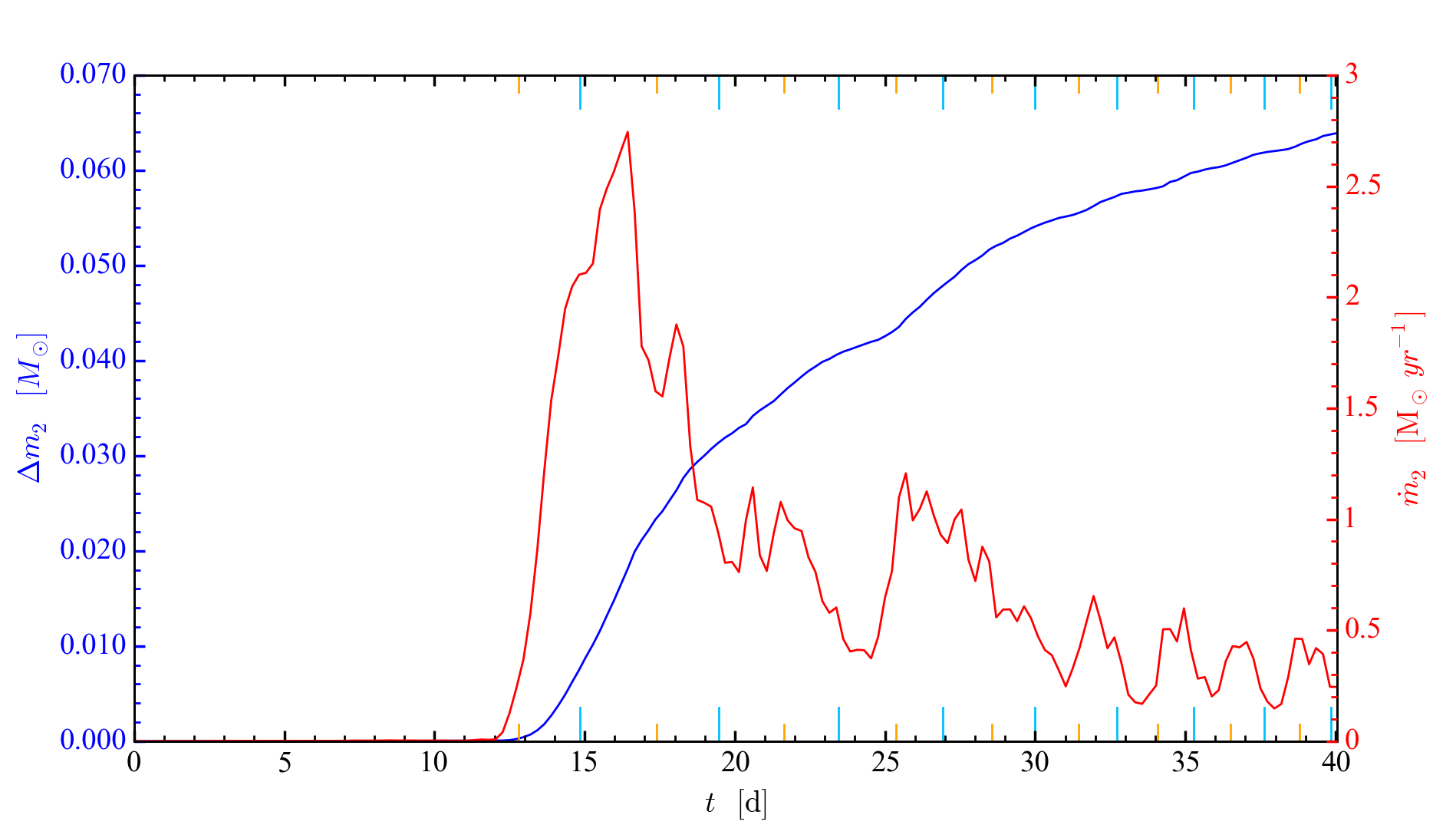} 
  \caption{\textbf{Top panel:} `Accretion' by the companion for Model~A, which is not true accretion because no subgrid accretion model is used.
           The total mass contained within spheres of various radii is shown in blue, with line styles corresponding to control radii (see legend).
           A vertical green line marks the time at which the softening length is reduced by half.
           \textbf{Bottom panel:} Accretion by the companion for Model~B, for which the \citetalias{Krumholz+04} subgrid accretion model is used.
           The accreted mass is shown in blue (left-hand vertical axis) and the accretion rate,
           obtained by differentiating the accreted mass, is shown in red (right-hand vertical axis).
           For both panels, long light blue (short orange) tick marks show the times of apastron (periastron) passage,
           seen in Figure~\ref{fig:orbit}.
           (Where the sampling rate for the inter-particle separation was lower, 
           times of apastron/periastron passage were obtained after performing an interpolation.)
           \label{fig:macc}
          }            
\end{figure*}

\begin{figure*}
  \includegraphics[height=75mm,clip=true,trim= 50 120 220 170]{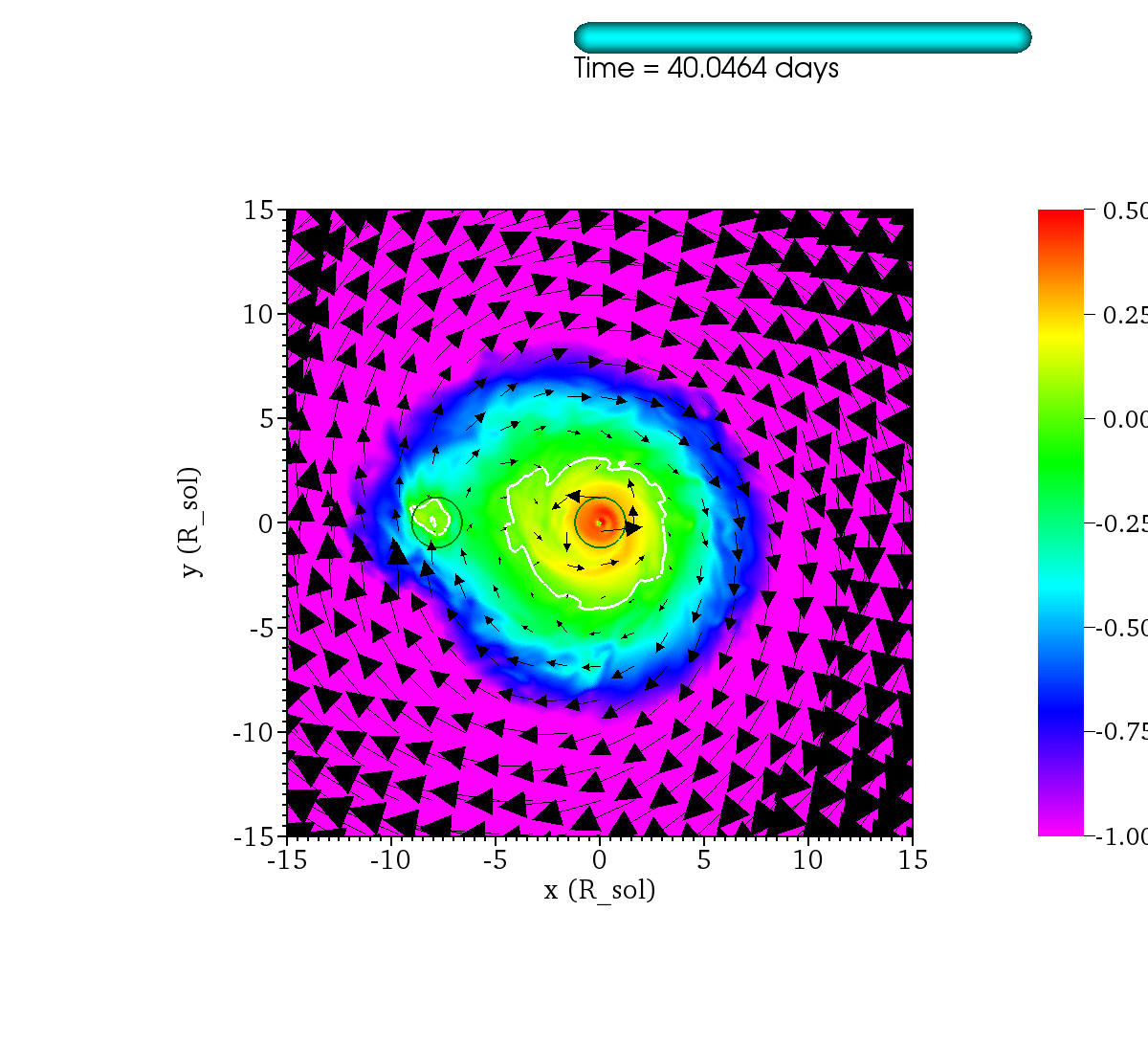}
  \includegraphics[height=75mm,clip=true,trim=220 120  40 170]{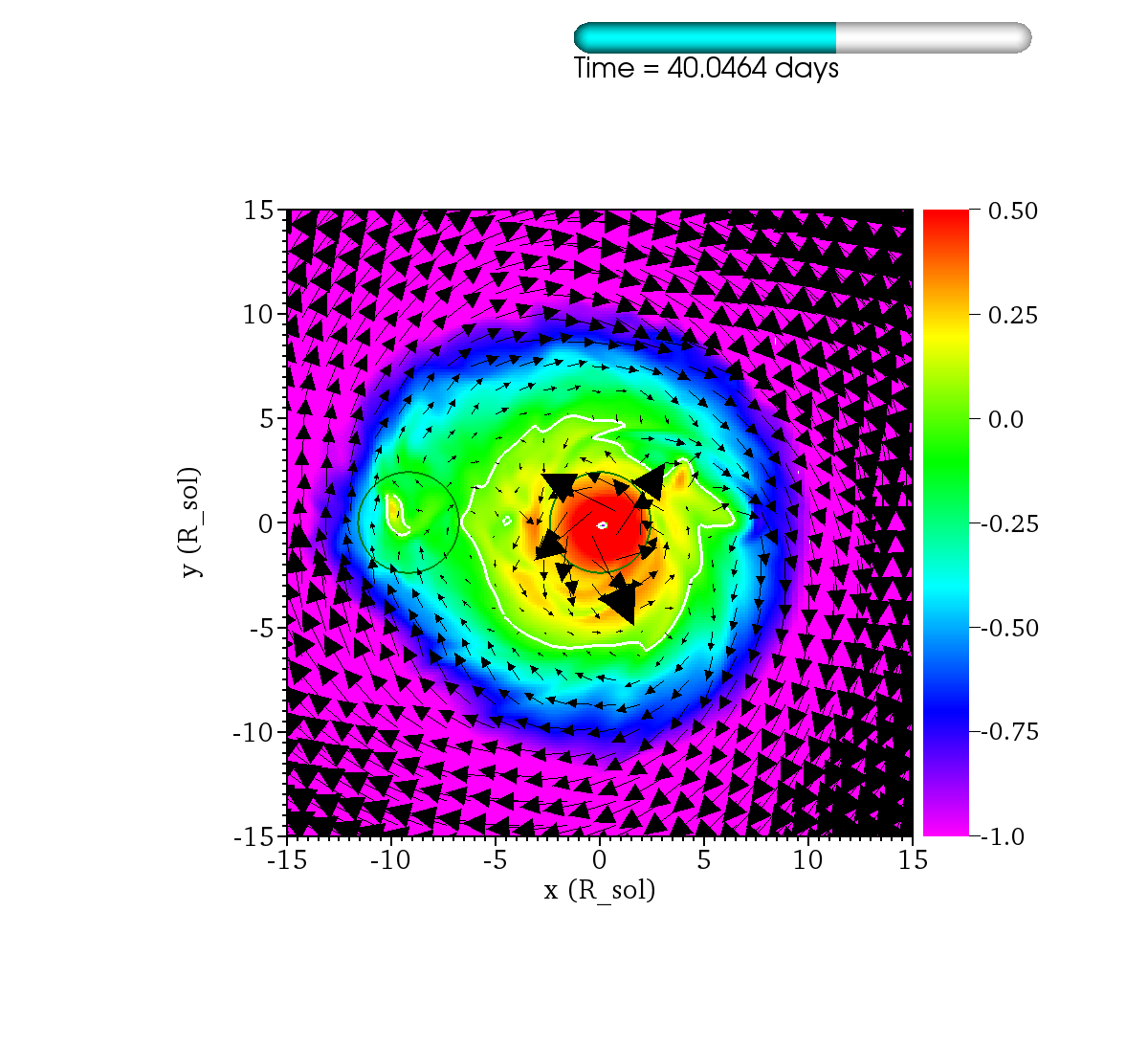}
  \caption{Slice through the orbital plane at $t=40\da$ with colour showing the tangential (with respect to the secondary, located at the centre of each panel)
           component of the velocity in the frame of reference rotating about the secondary with the instantaneous orbital angular velocity of the sink particles.
           Values are normalized by the local Keplerian circular speed around the secondary, corrected for the spline potential inside the softening sphere.
           The zero value, where the tangential component reverses direction, is shown by a white contour.
           Vectors show the direction and magnitude of the projection of this same velocity onto the orbital plane 
           (each vector refers to the location at which its tail begins).
           Model~A (no subgrid accretion) is shown on the left and Model~B (\citetalias{Krumholz+04} subgrid accretion) is shown on the right.
           Softening spheres are indicated by green circles.
           \label{fig:velocity}
          }
\end{figure*}

\subsection{`Accretion' in the absence of a subgrid model (Model A)}
\label{sec:accretion}

For Model~A, subgrid accretion is turned off, but we can measure the rate of mass flowing toward the secondary. 
We show this in the top panel of Figure~\ref{fig:macc} by calculating 
and plotting the gas mass contained inside spheres of a given `control' radius  centred on the secondary versus time \citepalias{Ricker+Taam08,Ricker+Taam12}.
The radius of the control sphere used for the different curves is shown in the legend.
The green vertical line indicates the time at which the softening length is halved. 
Times of apastron and periastron passage are marked on the horizontal axis with long blue and short orange tick-marks, respectively.
As stated earlier, the mass of the secondary and primary sink particles are $m_2=0.978M_\odot$, 
and $m_1=0.369M_\odot$ respectively.

The key result is that by the end of the simulation time, the mass flow toward the secondary has stopped. 
This is in sharp contrast to the case Model~B discussed in Section~\ref{sec:Krumholz_results}.

\subsubsection{More detailed description of time evolution}
The qualitative behaviour of the curves is approximately independent of the control radius
in that their shapes are very similar even though their amplitudes differ.
This tells us that the flow near the secondary is `global';  
different radii move inward or outward contemporaneously
(on average over each spherical control surface).

The inflow rate $\dot{M}\intwo$ is very small until $t\approx10\da$, increasing 
between $t\approx12.5\da$ and $t\approx13.5\da$, peaking later for smaller control radii.
During this time,  $M\intwo$ increases monotonically  before reaching a local maximum at $t=15\da$
and then decreases before increasing again (this  is most conspicuous for control radius $3R_\odot)$.
This local maximum coincides with the first maximum in the inter-particle separation curve of Figure~\ref{fig:orbit}.
As $M\intwo$  increases, the softening length is halved at $t=16.7\da$. 
This results in a prolonged increase, modulated by small oscillations, until $t\approx21\da$,  the time of the third periastron passage, The average interior mass then remains roughly constant, 
 with small oscillations.
The latter correspond to oscillations in the inter-particle separation,
with local maxima and minima of $M\intwo$ approximately coinciding with periastron and apastron passages respectively.
The mean value of $M\intwo$ slowly declines  after $t\approx28\da$.

The initial rise in $M\intwo$ is accompanied by a less pronounced rise in $M\inone$ until $t\approx13\da$ (just after the first periastron passage), 
followed by a sharp decrease (not shown).
Like $M\intwo$, $M\inone$ receives a `boost' immediately following the change in softening radius at $t=16.7\da$
followed by gradual decay, and modulated by oscillations that are approximately in phase with those of $M\intwo$.

These features can tentatively be explained as follows.
As the plunging-in secondary approaches the high-density RG core, it accretes at an ever higher rate,
until it has accreted a quasi-steady envelope.
The mean mass of this envelope over several orbits remains approximately constant.
The primary retains  part of the remnant RG envelope.
As the two particles approach, a larger portion of the primary envelope extends into the control sphere surrounding the secondary, 
increasing the integrated mass inside the control spheres around both particles.
When the particles separate, $M\intwo$ and $M\inone$ decrease again for the same reason. This back-and-forth  motion explains the aforementioned oscillations.

When the softening radius is reduced from $r\soft=r\softf$ to $r\soft=r\softf/2$, 
the depth of the potential well doubles at $r=0$ and the gravitational acceleration of each particle increases everywhere 
within the sphere of the original softening radius $r\softf$ centred on the particle.
Gas then flows toward the secondary until a more massive, more concentrated quasi-steady envelope establishes.
A weaker similar effect occurs for the less massive primary particle.  
The gradual decrease in the orbital separation is likely caused 
by gas dynamical friction on the secondary
(\citetalias{Ricker+Taam08,Ricker+Taam12}; \citealt{Macleod+17}).
This drag may be enhanced as the quasi-steady envelope around the secondary becomes more concentrated, thereby explaining the reduction in  mean separation $a$,
and, from Kepler's law, the reduction in orbital period.
There is opportunity to explore the dynamics in detail in future work.

Finally, the slow decrease in the secondary and primary interior masses
$M\intwo$ and $M\inone$
during the final $\approx12\da$ 
can tentatively be explained by the reduction in size of the Roche lobes as the inter-particle separation becomes smaller.

At the end of the simulation the mass flow toward the secondary stops.
The total mass `accreted' by the companion $M\intwo$ 
is $4\times10^{-4}$, $2\times10^{-3}$, $7\times10^{-3}$ and $1.3\times10^{-2}M_\odot$
for control radii $R_\odot/2$, $R_\odot$, $2R_\odot$ and $3R_\odot$, respectively.
The corresponding peak inflow rates $\dot{M}\intwo$ are $0.1$, $0.3$, $0.8$ and $1.9 M_\odot \yr^{-1}$, respectively.
The values of $M\intwo$ are similar to those of \citet{Ricker+Taam08,Ricker+Taam12},
though at the end of their simulation $M\intwo$ is still increasing.
However, their simulation lasted for a smaller number of orbits, 
so may only correspond to the early stages of our simulation before infall stops.

\subsection{`Accretion' with a subgrid model (Model B)}
\label{sec:Krumholz_results}
We now turn to Model~B, which includes \citetalias{Krumholz+04} subgrid accretion.
The key results are shown in the bottom panel of 
Figure~\ref{fig:macc}.

The pressure release valve provided by the Krumholz subgrid model in Model~B allows  mass to flow continuously onto the secondary without stopping, 
unlike  Model~A, 
for which the mass inflow stops and the envelope mass around the secondary remains quasi-steady.
The actual (subgrid) accretion rate onto the secondary is shown in the bottom panel of Figure~\ref{fig:macc}, 
where we plot the evolution of the change in secondary mass $\Delta m_2$, 
along with the rate of change $\dot{m}_2$.%
\footnote{The rate is calculated using a central difference method accurate to second order. 
The sampling rate is constant and approximately equal to one frame every $0.23\da$.}
Accretion begins at $t\approx12\da$, coinciding with the first periastron passage. 
The accretion rate $\dot{m}_2$ peaks between $t=16\da$ and $t=17\da$ at $\dot{m}_2\approx 2.7M_\odot\, \mathrm{yr}^{-1}$.
By the end of the simulation, the accretion rate reaches a  fairly steady value of $\dot{m}_2\approx 0.3M_\odot\, \mathrm{yr}^{-1}$, 
modulated by oscillations likely related to the oscillations in the inter-particle separation.
By  the end of the simulation at $t=40\da$ the secondary has accreted $0.064M_\odot$,
for a $6.5\%$ gain in mass, and continues to accrete.

The plots of $M\intwo$ and $M\inone$ and their rates of change for Model~B (not shown) are qualitatively similar to
those for Model~A, but  the fractional decrease in $M\intwo$ between its peak and $t=40\da$ 
is $\approx5$ times greater in Model~B than in Model~A, 
while the fractional decrease in $M\inone$ from the time that $M\intwo$ peaks to $t=40\da$ 
is only $\approx20$ per cent greater.
Since the main difference between Model~A and Model~B is the presence or absence of subgrid accretion,
this result is consistent with the finding that the gas flow around the secondary is influenced by the subgrid accretion in Model~B.%
\footnote{However, the reduction in softening length applied in Model~A but not in Model~B and the fact that in Model~B, 
$M\intwo$ peaks at $t\approx17\da$, much earlier than in Model~A, may also play a role.}

\subsection{Velocity field around the secondary}
\label{sec:flow}
Exploring the flow properties around the secondary helps to assess the plausibility of disc
formation at even smaller, unresolved distances from the secondary ($<r\soft$). 
In Figure~\ref{fig:velocity} we plot in color the local tangential velocity $v_{\phi,2}$
with respect to the secondary in the frame of reference that rotates at the instantaneous angular velocity of the particle orbit about the secondary.
Shown is a slice through the secondary perpendicular to the $z$-axis at $t=40\da$,
with green circles demarking the spline softening spheres.
Here $v_{\phi,2}$ is normalized with respect to the Keplerian speed $v\Kep$ about the secondary, corrected for the spline potential within the softening radius.
A white contour delineates where $v_{\phi,2}=0$, and the vectors show the direction and relative magnitude of the velocity
field of the gas projected onto the slice.

Figure~\ref{fig:velocity} shows that $v_{\phi,2}>0$ within about $3$ to $4R_\odot$ of the secondary for Model~A,
and within about $5$ to $6R_\odot$ of the secondary for Model~B.
Outside of this region, the gas rotates clockwise ($v_{\phi,2}<0$),  lagging the orbital motion of the particles in the lab frame.
Inside the region of counter-clockwise rotation, the vectors show that the gas exhibits both positive and negative radial velocity  $v_{r,2}$.
The magnitude of the tangential component is $\sim\tfrac{1}{5}V\Kep$ to $\sim\tfrac{1}{4}V\Kep$ 
within about $2R_\odot$ from the softening radius in both simulations.
We discuss the implications of this angular motion 
in the next section.

\section{Implications for accretion in CEE}
\label{sec:discussion}

\subsection{Condition for accretion disc formation}

The sub-Keplerian speeds of the gas flowing around the secondary imply that a  relatively thick  disc forms on resolved scales, 
although this does not by itself determine the disc structure  within $r<r\soft$ from the secondary,
where gravity is not treated fully realistically in our model (see also the next subsection).
This point is particularly germane when  the secondary is a WD  with radius $R_*\sim0.01R_\odot$--
far smaller than the softening radius and comparable to the size of the smallest resolution element.

To assess the minimum condition needed for an inner thin disc to form, we can simply
assume  conservation of specific angular momentum of 
the flow  around the secondary at $r\gtrsim r\soft$ 
and determine if the flow would be purely rotationally supported at $r\gtrsim R_*$.
Assuming that any net angular momentum transport would be outward,
 this condition sets an upper limit on the stellar radius that would still leave room for a Keplerian thin disc to form.
The specific angular momentum at radius $r=r\soft$ (averaged over azimuth, in the orbital plane) can be written as $f\sqrt{\Gn m_2 r\soft}$,
where $\sqrt{\Gn m_2 r}$ is the Keplerian value at radius $r$ and $f\leq1$ is a parameter that can be estimated from the simulations.
The upper limit to the stellar radius for disc formation is then given by
\begin{equation}
  R_* \lesssim f^2r\soft.
\end{equation}

Section~\ref{sec:flow} shows that $1/5\lesssim f\lesssim 1/4$ for both Model~A ($r\soft(40\da)=1.2R_\odot$) and Model~B ($r\soft=2.4R_\odot$). 
This translates into a maximum stellar radius of $0.05R_\odot\lesssim R\starmax\lesssim 0.08R_\odot$ for Model~A and 
$0.10R_\odot\lesssim R\starmax\lesssim 0.15R_\odot$
for Model~B.
This result suggests that a purely Keplerian thin accretion disc would not be able to form  if the star is a MS star ($R_*\sim1R_\odot$), 
but that there is still plenty of room for a thin accretion disc to form if the secondary is a WD ($R_*\sim0.01R_\odot$).
Disc thinness is not a necessary property for dynamically significant accretion and a jet. 
Super-Eddington accretion could produce a radiation-dominated thick disc which facilitates collimated radiation-driven outflows. 
However, for jets that depend on Poynting flux anchored in the disc, 
the outflow power is proportional to the angular velocity at the field anchor point so too slow a rotator would reduce the outflow power. 
All of this warrants further work.

\subsection{Need for a pressure valve}
Comparing the no-accretion case with the \citetalias{Krumholz+04} case highlights that sustained long term accretion requires a pressure valve. 
The \citetalias{Krumholz+04} prescription takes away both mass and pressure, allowing material to continue to infall at the inner boundary.  
If we disallow this infall, the accretion flow eventually ceases.  
The \citetalias{Krumholz+04} prescription was originally developed for protostellar accretion where material 
that accretes onto the central object can lose its pressure via radiation through optically thin gas. 
In the present case, the gas is optically thick, so we do not expect such pressure to release radiation.  
Jets \citep{Staff+16,Morenomendez+17,Soker17b} provide a more likely alternative.  
In fact, we suspect that jets may provide the {\it only} way to sustain accretion in these dense environments, 
implying a mutually symbiotic relation and one-to-one correspondence between the two.  
Since a  jet would provide a pressure release valve that is not as large as that of the \citetalias{Krumholz+04} prescription,  
our two cases (no accretion vs. accretion)  bound the  extreme cases of maximum and zero valve.

\subsection{Super-Eddington Accretion}
As material accretes from the envelope to the secondary,
we estimate from the virial theorem that about half of the liberated potential energy will ultimately be added to the accretor in the form of thermal energy.
The remainder will be transferred to the optically thick envelope
in the form of bulk kinetic and thermal energy.
The rate of this energy transfer can be estimated as
\begin{equation}
  \label{Edot_acc}
  \dot{E}\acc= \frac{\Gn m_2 \dot{m}_2}{2\lamtilde R_*},
\end{equation}
where $R_*$ is the radius of the accretor,
and $\lamtilde$ is a parameter of order unity that depends on the density profile of the accretor 
and is analogous to the parameter $\lambda$ often invoked for the primary envelope in CE calculations \citep[][and references therein]{Ivanova+13}.
For this formula, we assumed that $1/R_*\gg 1/r\init$,
where $r\init$ is the initial radius of the accreting gas with respect to the centre of the accretor.
We have also neglected the change in potential energy associated with the primary particle and with the self-gravity of gas.

Assuming an opacity  due to electron scattering, 
the Eddington accretion rate onto the secondary is given by
\begin{equation}
  \label{m2_Edd}
  \dot{m}_{2,\,\mathrm{Edd}}= \frac{4\pi m\proton\Gn m_2}{\epsilon c\sigma_\mathrm{T}},
\end{equation}
where $m\proton$ is the proton mass, $c$ is the speed of light in vacuum,
$\sigma_\mathrm{T}$ is the Thomson scattering cross section, and $\sigma_\mathrm{T}/m\proton=0.398\cm^2\g^{-1}$.
The radiative efficiency $\epsilon$ is the fraction of the rest-mass energy of accreted material 
that is transferred to the envelope.
Thus we write
\begin{equation}
  \label{epsilon}
  \epsilon= \frac{\dot{E}\acc}{\dot{m}_2 c^2},
\end{equation}
where $\dot{E}\acc$ is the rate of energy transferred to the envelope due to accretion.
Substituting expression~\eqref{Edot_acc} into equation~\eqref{epsilon} and then the resulting expression for $\epsilon$ into equation~\eqref{m2_Edd}, 
we obtain an efficiency
\begin{equation}
  \epsilon= 1.1\times10^{-6}\;\lamtilde^{-1}\left(\frac{m_2}{M_\odot}\right)\left(\frac{R_*}{R_\odot}\right)^{-1}
\end{equation}
and an Eddington accretion rate
\begin{equation}
  \dot{m}_{2,\,\mathrm{Edd}}= 2.1\times10^{-3}\Msol\yr^{-1}\;\lamtilde\left(\frac{R_*}{R_\odot}\right).
\end{equation}
This is not a strict upper limit because it assumes energy release as radiation and spherical symmetry,
so can be circumvented to some extent if the energy is transported by a bipolar jet.
From the bottom panel of Figure~\ref{fig:macc}, we see that accretion rates of order $0.2$ to $2\Msol\yr^{-1}$ are obtained for our $\sim1M_\odot$ secondary.
For a MS star with radius $1\Rsol$, this translates to accretion rates  $10^2$ to $10^3$ times larger than the Eddington rate,
while for a WD with radius $0.01\Rsol$, we find that $\dot{m}_2$ is $10^4$ to $10^5$ times the Eddington rate.

\subsection{Implications for CE  PNe and PPNe}

The possibility of so-called super-Eddington or hypercritical accretion and its potential role in unbinding the CE is discussed in \citet{Ivanova+13,Shiber+16}.
Although steady  accretion onto a WD or MS star at such extreme rates
would likely be substantially mitigated by feedback from the accretion processes closer to the stellar surface, 
super-Eddington values may be possible.
 
As discussed in the introduction jets produced by accretion also play an important role in shaping PPNe and PNe.
Depending on the engine (WD or MS star), momentum requirements for a number of PPNe would seem to require
accretion rates close to this limit and would have to be achieved  either within the CE or the 
Roche-lobe overflow phase just before \cite{Blackman+2014}.
Were it the case that a jet could not form or sustain inside the CE because the pressure valve were simply not efficient enough, 
we would be led to conclude that evidence for accretion-powered outflows in PPNe could not emanate from inside the CE 
but more likely from the Roche Lobe overflow phase \citep{Staff+16} before CE.
There is much  opportunity for future work to study the  fate of a jet formed outside the CE as it enters the CE.

\section{Numerical Challenges and Need for Convergence Study}
\label{sec:convergence}
In  Section~\ref{sec:results} we explained how the particles' orbits and the flow around the secondary are affected when the spline softening length $r\soft$ is halved
in Model~A from $2.4R_\odot$ to $1.2R_\odot$ at $t=16.7\da$.
We argued that this change leads to a reduced orbital period and separation, and to a denser envelope of material around the secondary
than what would have obtained had $r\soft$ been kept constant as in Model~B.
Such behaviour is unphysical because the gravitational softening is a numerical device that is imposed to deal with finite resolution.
In reality, there would be a stellar surface, at $r\sim1R_\odot$ for a MS secondary or at $r\sim0.01R_\odot$ for a WD secondary.
For the latter,  any imposed softening length should ideally be $\ll1R_\odot$, and yet the softening length must be adequately resolved,
e.g.~to ensure energy conservation \citepalias{Ohlmann+16a}.

In Model~A we followed \citetalias{Ohlmann+16a} by requiring that $r\soft$ not exceed $1/5$ of the inter-particle separation.
That this  changes the orbit and flow  suddenly, when the softening length is halved just before the threshold $1/5$ is reached, 
implies that $1/5$ is too large.
This suggests  that finite softening lengths produce overestimates of the inter-particle separations  at late times in global CE simulations.
This is consistent with a conclusion of \citet{Iaconi+17b}, 
that final separations  decrease with decreasing smoothing length.

What makes CE simulations computationally demanding is the need to  simultaneously satisfy three crucial constraints.
First is that the resolution near the particles should be high enough \citep{Iaconi+17b}, particularly within the  softening length \citepalias{Ohlmann+16a}.
The resolution can affect the orbit and fidelity of energy conservation in the final phases of inspiral.
Second, the softening length itself must be sufficiently small, for the reasons discussed above.
But  making $r\soft$ comparable to the stellar radius is also problematic because, as we have emphasized, 
the choice of subgrid model for physics near the stellar surface (e.g.~accretion, jets, convection) becomes influential.
Third, the spatial extent of the region surrounding the particles within which the resolution is highest must be large enough
to prevent inaccuracies in the particles' orbit for instance. This can arise due to the particles' non-circular orbit and the complex nature of the flow:  particles can end up interacting strongly with gas that had not been well-resolved at some earlier time.

Other  choices about the numerical setup must be made, like the size of the domain and base resolution,
but the three considerations listed above are  most crucial for studying CEE near the particles.

In this work, we chose values for the size of the smallest resolution element ($0.07$ to $0.14R_\odot$), 
softening length ($1.2$ to $2.4R_\odot$), and refinement radius (see Figure~\ref{fig:orbit}) 
that are `conservative'  to  ensure qualitatively physical results.
However,  a convergence study that explores each of these parameters separately
while keeping the other two  fixed, is warranted.
But given the extensive computational resources required,
we leave such a  comprehensive study for future work.

Preliminary tests at lower resolution do suggest that each of the three constraints can be important,
and we present an analysis of a limited set of test runs in Appendix~\ref{sec:test_runs}.
Depending on the goals of a particular investigation,
a convergence study can help determine which one or more of the three parameter values can be relaxed 
(i.e.~our choices could be adjusted to be more `aggressive'), and in turn, help optimize the use of computational resources. 

\section{Conclusions}
\label{sec:conclusions}

We have presented a new platform for simulating CEE.  
Our AstroBEAR AMR MHD multi-physics code \citep{Cunningham+09,Carroll-nellenback+12} 
has been adapted to model the interaction of an extended giant star with a companion modeled as a point mass. 
We have used the platform to carry out two high resolution simulations of CE interaction with the goal 
of assessing the nature of accretion onto the secondary.  
Our Model~B employed a subgrid model for accretion that removes mass and pressure from the grid, 
effectively making the secondary a point mass ``sink'' particle \citepalias{Krumholz+04}.
No subgrid model was employed in our Model A simulation 
and so gravitationally bound material that collected around the secondary was not removed from the grid. 

For both simulations, AstroBEAR accurately models the main features of CEE when compared to previous simulations.  
In particular, the code captures the rapid inspiral of the secondary as well as the conservation of orbital energy 
into mass motions of the envelope which can unbind a fraction of the gas 

With regard to our main focus question of accretion onto the secondary,  
we find that only for Model~B, where  a subgrid model is used to remove mass and pressure near the secondary particle, 
does gas continue to accrete onto the secondary core. 
In contrast, for  Model~A, mass collects around the secondary, 
forming a quasi-stationary extended high pressure atmosphere whose accretion eventually shuts off. 
This is  important  because accretion and accretion discs are tied to the generation of outflows/jets. 
Such outflows from deep in a stellar interior have been considered as the means for driving some classes of supernova \citep{Milosavljevic+12,Gilkis+16}. 
As \citet{Sabach+Soker15,Soker15,Shiber+17,Shiber+Soker18} have shown in their ``Grazing Envelope'' models, 
the presence of strong jets can lead to the envelope material being driven out of the CE system. 
We have shown that accretion onto the secondary in CEE can produce super-Eddington accretion rates on either a MS star or WD 
but only if there is a ``pressure release valve'' for maintaining steady flows during the inspiral. 
Since the CE environments are optically thick we are led to speculate that jets may be the only way to sustain accretion. 
As such, there would be a one-to-one correspondence between active accretion and jets in CEE. 
 
If bipolar outflows can be sustained in CEE, they are candidates to supply the outflow momentum and energy budgets needed to explain PPN bipolar jets. 
Otherwise, accretion-powered jets could still be produced in the Roche lobe phase just before entering the CE.

Finally, as was noted in the introduction, the ejection of the envelope and the completion of the inspiral to small radii 
has proven difficult to achieve  in simulations \citep{Ohlmann+16a,Kuruwita+16} including ours without additional feedback.  
Explanations for this behaviour split between those focusing on limits of the {\it numerics} and those that focus on limits in the {\it physics}. 
For numerics the concern relates to resolution and timescales.  
For physics the concern is that there are important physical processes not included in the numerical models.  
Examples of such processes include recombination and radiation pressure on dust. 
While recombination has received positive attention, 
newer work casts some doubt on its efficacy \citep{Soker+Harpaz03,Sabach+17,Grichener+18}.
There is much opportunity to further delineate the role of accretion and outflows in CEE alongside the other envelope loss physics mechanisms, 
and the mutual connection if any, to further orbital decay.

\section*{Acknowledgements}
LC wishes to thank Sebastian~Ohlmann for helpful discussions relating to methods.
The authors gratefully acknowledge
Orsola~De~Marco, Paul~Ricker, Morgan~MacLeod, Brian~Metzger, Natalia~Ivanova, Noam~Soker and Hui~Li
for thought-provoking conversations during the time this work was being prepared.  
JN acknowledges financial support from NASA grants HST-15044 and HST-14563.

\footnotesize{
\noindent
\bibliographystyle{mnras}
\bibliography{refs}

\begin{thebibliography}{}
\makeatletter
\relax
\def\mn@urlcharsother{\let\do\@makeother \do\$\do\&\do\#\do\^\do\_\do\%\do\~}
\def\mn@doi{\begingroup\mn@urlcharsother \@ifnextchar [ {\mn@doi@}
  {\mn@doi@[]}}
\def\mn@doi@[#1]#2{\def\@tempa{#1}\ifx\@tempa\@empty \href
  {http://dx.doi.org/#2} {doi:#2}\else \href {http://dx.doi.org/#2} {#1}\fi
  \endgroup}
\def\mn@eprint#1#2{\mn@eprint@#1:#2::\@nil}
\def\mn@eprint@arXiv#1{\href {http://arxiv.org/abs/#1} {{\tt arXiv:#1}}}
\def\mn@eprint@dblp#1{\href {http://dblp.uni-trier.de/rec/bibtex/#1.xml}
  {dblp:#1}}
\def\mn@eprint@#1:#2:#3:#4\@nil{\def\@tempa {#1}\def\@tempb {#2}\def\@tempc
  {#3}\ifx \@tempc \@empty \let \@tempc \@tempb \let \@tempb \@tempa \fi \ifx
  \@tempb \@empty \def\@tempb {arXiv}\fi \@ifundefined
  {mn@eprint@\@tempb}{\@tempb:\@tempc}{\expandafter \expandafter \csname
  mn@eprint@\@tempb\endcsname \expandafter{\@tempc}}}

\bibitem[\protect\citeauthoryear{{Abbott} et~al.,}{{Abbott}
  et~al.}{2016}]{Abbott+16}
{Abbott} B.~P.,  et~al., 2016, \mn@doi [Physical Review Letters]
  {10.1103/PhysRevLett.116.061102}, \href
  {http://adsabs.harvard.edu/abs/2016PhRvL.116f1102A} {116, 061102}

\bibitem[\protect\citeauthoryear{{Abbott} et~al.,}{{Abbott}
  et~al.}{2017a}]{Abbott+17a}
{Abbott} B.~P.,  et~al., 2017a, \mn@doi [Physical Review Letters]
  {10.1103/PhysRevLett.119.141101}, \href
  {http://adsabs.harvard.edu/abs/2017PhRvL.119n1101A} {119, 141101}

\bibitem[\protect\citeauthoryear{{Abbott} et~al.,}{{Abbott}
  et~al.}{2017b}]{Abbott+17b}
{Abbott} B.~P.,  et~al., 2017b, \mn@doi [Physical Review Letters]
  {10.1103/PhysRevLett.119.161101}, \href
  {http://adsabs.harvard.edu/abs/2017PhRvL.119p1101A} {119, 161101}

\bibitem[\protect\citeauthoryear{{Balick} \& {Frank}}{{Balick} \&
  {Frank}}{2002}]{Balick+Frank02}
{Balick} B.,  {Frank} A.,  2002, \mn@doi [\araa]
  {10.1146/annurev.astro.40.060401.093849}, \href
  {http://adsabs.harvard.edu/abs/2002ARA%26A..40..439B} {40, 439}

\bibitem[\protect\citeauthoryear{{Belczynski}, {Buonanno}, {Cantiello},
  {Fryer}, {Holz}, {Mandel}, {Miller}  \& {Walczak}}{{Belczynski}
  et~al.}{2014}]{Belczynski+14}
{Belczynski} K.,  {Buonanno} A.,  {Cantiello} M.,  {Fryer} C.~L.,  {Holz}
  D.~E.,  {Mandel} I.,  {Miller} M.~C.,   {Walczak} M.,  2014, \mn@doi [\apj]
  {10.1088/0004-637X/789/2/120}, \href
  {http://adsabs.harvard.edu/abs/2014ApJ...789..120B} {789, 120}

\bibitem[\protect\citeauthoryear{{Blackman} \& {Lucchini}}{{Blackman} \&
  {Lucchini}}{2014}]{Blackman+2014}
{Blackman} E.~G.,  {Lucchini} S.,  2014, \mn@doi [\mnras]
  {10.1093/mnrasl/slu001}, \href
  {http://adsabs.harvard.edu/abs/2014MNRAS.440L..16B} {440, L16}

\bibitem[\protect\citeauthoryear{{Blackman}, {Frank}  \& {Welch}}{{Blackman}
  et~al.}{2001}]{Blackman+2001}
{Blackman} E.~G.,  {Frank} A.,   {Welch} C.,  2001, \mn@doi [\apj]
  {10.1086/318253}, \href {http://adsabs.harvard.edu/abs/2001ApJ...546..288B}
  {546, 288}

\bibitem[\protect\citeauthoryear{{Blank}, {Morris}, {Frank},
  {Carroll-Nellenback}  \& {Duschl}}{{Blank} et~al.}{2016}]{Blank+16}
{Blank} M.,  {Morris} M.~R.,  {Frank} A.,  {Carroll-Nellenback} J.~J.,
  {Duschl} W.~J.,  2016, \mn@doi [\mnras] {10.1093/mnras/stw771}, \href
  {http://adsabs.harvard.edu/abs/2016MNRAS.459.1721B} {459, 1721}

\bibitem[\protect\citeauthoryear{{Bondi}}{{Bondi}}{1952}]{Bondi52}
{Bondi} H.,  1952, \mn@doi [\mnras] {10.1093/mnras/112.2.195}, \href
  {http://adsabs.harvard.edu/abs/1952MNRAS.112..195B} {112, 195}

\bibitem[\protect\citeauthoryear{{Bondi} \& {Hoyle}}{{Bondi} \&
  {Hoyle}}{1944}]{Bondi+Hoyle44}
{Bondi} H.,  {Hoyle} F.,  1944, \mn@doi [\mnras] {10.1093/mnras/104.5.273},
  \href {http://adsabs.harvard.edu/abs/1944MNRAS.104..273B} {104, 273}

\bibitem[\protect\citeauthoryear{{Bujarrabal}, {Castro-Carrizo}, {Alcolea}  \&
  {S{\'a}nchez Contreras}}{{Bujarrabal} et~al.}{2001}]{Bujarrabal+01}
{Bujarrabal} V.,  {Castro-Carrizo} A.,  {Alcolea} J.,   {S{\'a}nchez Contreras}
  C.,  2001, \mn@doi [\aap] {10.1051/0004-6361:20011090}, \href
  {http://adsabs.harvard.edu/abs/2001A%26A...377..868B} {377, 868}

\bibitem[\protect\citeauthoryear{{Carroll-Nellenback}, {Shroyer}, {Frank}  \&
  {Ding}}{{Carroll-Nellenback} et~al.}{2012}]{Carroll-nellenback+12}
{Carroll-Nellenback} J.,  {Shroyer} B.,  {Frank} A.,   {Ding} C.,  2012, in
  {Pogorelov} N.~V.,  {Font} J.~A.,  {Audit} E.,   {Zank} G.~P.,  eds,
  Astronomical Society of the Pacific Conference Series Vol. 459, Numerical
  Modeling of Space Plasma Slows (ASTRONUM 2011). p.~291 (\mn@eprint {arXiv}
  {1112.1710})

\bibitem[\protect\citeauthoryear{{Carroll-Nellenback}, {Shroyer}, {Frank}  \&
  {Ding}}{{Carroll-Nellenback} et~al.}{2013}]{Carroll-nellenback+13}
{Carroll-Nellenback} J.~J.,  {Shroyer} B.,  {Frank} A.,   {Ding} C.,  2013,
  \mn@doi [Journal of Computational Physics] {10.1016/j.jcp.2012.10.004}, \href
  {http://adsabs.harvard.edu/abs/2013JCoPh.236..461C} {236, 461}

\bibitem[\protect\citeauthoryear{{Cunningham}, {Frank}, {Varni{\`e}re},
  {Mitran}  \& {Jones}}{{Cunningham} et~al.}{2009}]{Cunningham+09}
{Cunningham} A.~J.,  {Frank} A.,  {Varni{\`e}re} P.,  {Mitran} S.,   {Jones}
  T.~W.,  2009, \mn@doi [\apjs] {10.1088/0067-0049/182/2/519}, \href
  {http://adsabs.harvard.edu/abs/2009ApJS..182..519C} {182, 519}

\bibitem[\protect\citeauthoryear{{De Marco} \& {Izzard}}{{De Marco} \&
  {Izzard}}{2017}]{Demarco+Izzard17}
{De Marco} O.,  {Izzard} R.~G.,  2017, \mn@doi [\pasa] {10.1017/pasa.2016.52},
  \href {http://adsabs.harvard.edu/abs/2017PASA...34....1D} {34, e001}

\bibitem[\protect\citeauthoryear{{Edgar}}{{Edgar}}{2004}]{Edgar04}
{Edgar} R.,  2004, \mn@doi [\nar] {10.1016/j.newar.2004.06.001}, \href
  {http://adsabs.harvard.edu/abs/2004NewAR..48..843E} {48, 843}

\bibitem[\protect\citeauthoryear{{Eggleton}}{{Eggleton}}{1983}]{Eggleton83}
{Eggleton} P.~P.,  1983, \mn@doi [\apj] {10.1086/160960}, \href
  {http://adsabs.harvard.edu/abs/1983ApJ...268..368E} {268, 368}

\bibitem[\protect\citeauthoryear{{Frank}, {King}  \& {Raine}}{{Frank}
  et~al.}{2002}]{Frank+02}
{Frank} J.,  {King} A.,   {Raine} D.~J.,  2002, {Accretion Power in
  Astrophysics: Third Edition}

\bibitem[\protect\citeauthoryear{{Gilkis}, {Soker}  \& {Papish}}{{Gilkis}
  et~al.}{2016}]{Gilkis+16}
{Gilkis} A.,  {Soker} N.,   {Papish} O.,  2016, \mn@doi [\apj]
  {10.3847/0004-637X/826/2/178}, \href
  {http://adsabs.harvard.edu/abs/2016ApJ...826..178G} {826, 178}

\bibitem[\protect\citeauthoryear{{Glanz} \& {Perets}}{{Glanz} \&
  {Perets}}{2018}]{Glanz+Perets18}
{Glanz} H.,  {Perets} H.~B.,  2018, preprint, \href
  {http://adsabs.harvard.edu/abs/2018arXiv180108130G} {} (\mn@eprint {arXiv}
  {1801.08130})

\bibitem[\protect\citeauthoryear{{Grichener}, {Sabach}  \& {Soker}}{{Grichener}
  et~al.}{2018}]{Grichener+18}
{Grichener} A.,  {Sabach} E.,   {Soker} N.,  2018, preprint, \href
  {http://adsabs.harvard.edu/abs/2018arXiv180305864G} {} (\mn@eprint {arXiv}
  {1803.05864})

\bibitem[\protect\citeauthoryear{{Hoyle} \& {Lyttleton}}{{Hoyle} \&
  {Lyttleton}}{1939}]{Hoyle+Lyttleton39}
{Hoyle} F.,  {Lyttleton} R.~A.,  1939, \mn@doi [Proceedings of the Cambridge
  Philosophical Society] {10.1017/S0305004100021150}, \href
  {http://adsabs.harvard.edu/abs/1939PCPS...35..405H} {35, 405}

\bibitem[\protect\citeauthoryear{{Iaconi}, {Reichardt}, {Staff}, {De Marco},
  {Passy}, {Price}, {Wurster}  \& {Herwig}}{{Iaconi} et~al.}{2017}]{Iaconi+17a}
{Iaconi} R.,  {Reichardt} T.,  {Staff} J.,  {De Marco} O.,  {Passy} J.-C.,
  {Price} D.,  {Wurster} J.,   {Herwig} F.,  2017, \mn@doi [\mnras]
  {10.1093/mnras/stw2377}, \href
  {http://adsabs.harvard.edu/abs/2017MNRAS.464.4028I} {464, 4028}

\bibitem[\protect\citeauthoryear{{Iaconi}, {De Marco}, {Passy}  \&
  {Staff}}{{Iaconi} et~al.}{2018}]{Iaconi+17b}
{Iaconi} R.,  {De Marco} O.,  {Passy} J.-C.,   {Staff} J.,  2018, \mn@doi
  [\mnras] {10.1093/mnras/sty794}, \href
  {http://adsabs.harvard.edu/abs/2018MNRAS.tmp..774I} {}

\bibitem[\protect\citeauthoryear{{Iben} \& {Livio}}{{Iben} \&
  {Livio}}{1993}]{Iben+Livio93}
{Iben} Jr. I.,  {Livio} M.,  1993, \mn@doi [\pasp] {10.1086/133321}, \href
  {http://adsabs.harvard.edu/abs/1993PASP..105.1373I} {105, 1373}

\bibitem[\protect\citeauthoryear{{Ivanova} \& {Nandez}}{{Ivanova} \&
  {Nandez}}{2016}]{Ivanova+Nandez16}
{Ivanova} N.,  {Nandez} J.~L.~A.,  2016, \mn@doi [\mnras]
  {10.1093/mnras/stw1676}, \href
  {http://adsabs.harvard.edu/abs/2016MNRAS.462..362I} {462, 362}

\bibitem[\protect\citeauthoryear{{Ivanova} et~al.,}{{Ivanova}
  et~al.}{2013}]{Ivanova+13}
{Ivanova} N.,  et~al., 2013, \mn@doi [\aapr] {10.1007/s00159-013-0059-2}, \href
  {http://adsabs.harvard.edu/abs/2013A%26ARv..21...59I} {21, 59}

\bibitem[\protect\citeauthoryear{{Jones} \& {Boffin}}{{Jones} \&
  {Boffin}}{2017}]{Jones+Boffin17}
{Jones} D.,  {Boffin} H.~M.~J.,  2017, \mn@doi [Nature Astronomy]
  {10.1038/s41550-017-0117}, \href
  {http://adsabs.harvard.edu/abs/2017NatAs...1E.117J} {1, 0117}

\bibitem[\protect\citeauthoryear{{Kalogera}, {Belczynski}, {Kim},
  {O'Shaughnessy}  \& {Willems}}{{Kalogera} et~al.}{2007}]{Kalogera+07}
{Kalogera} V.,  {Belczynski} K.,  {Kim} C.,  {O'Shaughnessy} R.,   {Willems}
  B.,  2007, \mn@doi [\physrep] {10.1016/j.physrep.2007.02.008}, \href
  {http://adsabs.harvard.edu/abs/2007PhR...442...75K} {442, 75}

\bibitem[\protect\citeauthoryear{{Krumholz}, {McKee}  \& {Klein}}{{Krumholz}
  et~al.}{2004}]{Krumholz+04}
{Krumholz} M.~R.,  {McKee} C.~F.,   {Klein} R.~I.,  2004, \mn@doi [\apj]
  {10.1086/421935}, \href {http://adsabs.harvard.edu/abs/2004ApJ...611..399K}
  {611, 399}

\bibitem[\protect\citeauthoryear{{Kuruwita}, {Staff}  \& {De Marco}}{{Kuruwita}
  et~al.}{2016}]{Kuruwita+16}
{Kuruwita} R.~L.,  {Staff} J.,   {De Marco} O.,  2016, \mn@doi [\mnras]
  {10.1093/mnras/stw1414}, \href
  {http://adsabs.harvard.edu/abs/2016MNRAS.461..486K} {461, 486}

\bibitem[\protect\citeauthoryear{{Li}, {Frank}  \& {Blackman}}{{Li}
  et~al.}{2014}]{Li+14}
{Li} S.,  {Frank} A.,   {Blackman} E.~G.,  2014, \mn@doi [\mnras]
  {10.1093/mnras/stu1571}, \href
  {http://adsabs.harvard.edu/abs/2014MNRAS.444.2884L} {444, 2884}

\bibitem[\protect\citeauthoryear{{Livio} \& {Soker}}{{Livio} \&
  {Soker}}{1988}]{Livio+Soker88}
{Livio} M.,  {Soker} N.,  1988, \mn@doi [\apj] {10.1086/166419}, \href
  {http://adsabs.harvard.edu/abs/1988ApJ...329..764L} {329, 764}

\bibitem[\protect\citeauthoryear{{Lombardi}, {Proulx}, {Dooley}, {Theriault},
  {Ivanova}  \& {Rasio}}{{Lombardi} et~al.}{2006}]{Lombardi+06}
{Lombardi} Jr. J.~C.,  {Proulx} Z.~F.,  {Dooley} K.~L.,  {Theriault} E.~M.,
  {Ivanova} N.,   {Rasio} F.~A.,  2006, \mn@doi [\apj] {10.1086/499938}, \href
  {http://adsabs.harvard.edu/abs/2006ApJ...640..441L} {640, 441}

\bibitem[\protect\citeauthoryear{{MacLeod} \& {Ramirez-Ruiz}}{{MacLeod} \&
  {Ramirez-Ruiz}}{2015a}]{Macleod+Ramirez-ruiz15a}
{MacLeod} M.,  {Ramirez-Ruiz} E.,  2015a, \mn@doi [\apjl]
  {10.1088/2041-8205/798/1/L19}, \href
  {http://adsabs.harvard.edu/abs/2015ApJ...798L..19M} {798, L19}

\bibitem[\protect\citeauthoryear{{MacLeod} \& {Ramirez-Ruiz}}{{MacLeod} \&
  {Ramirez-Ruiz}}{2015b}]{Macleod+Ramirez-ruiz15b}
{MacLeod} M.,  {Ramirez-Ruiz} E.,  2015b, \mn@doi [\apj]
  {10.1088/0004-637X/803/1/41}, \href
  {http://adsabs.harvard.edu/abs/2015ApJ...803...41M} {803, 41}

\bibitem[\protect\citeauthoryear{{MacLeod}, {Antoni}, {Murguia-Berthier},
  {Macias}  \& {Ramirez-Ruiz}}{{MacLeod} et~al.}{2017}]{Macleod+17}
{MacLeod} M.,  {Antoni} A.,  {Murguia-Berthier} A.,  {Macias} P.,
  {Ramirez-Ruiz} E.,  2017, \mn@doi [\apj] {10.3847/1538-4357/aa6117}, \href
  {http://adsabs.harvard.edu/abs/2017ApJ...838...56M} {838, 56}

\bibitem[\protect\citeauthoryear{{MacLeod}, {Ostriker}  \& {Stone}}{{MacLeod}
  et~al.}{2018}]{Macleod+18}
{MacLeod} M.,  {Ostriker} E.~C.,   {Stone} J.~M.,  2018, preprint, \href
  {http://adsabs.harvard.edu/abs/2018arXiv180303261M} {} (\mn@eprint {arXiv}
  {1803.03261})

\bibitem[\protect\citeauthoryear{{Milosavljevi{\'c}}, {Lindner}, {Shen}  \&
  {Kumar}}{{Milosavljevi{\'c}} et~al.}{2012}]{Milosavljevic+12}
{Milosavljevi{\'c}} M.,  {Lindner} C.~C.,  {Shen} R.,   {Kumar} P.,  2012,
  \mn@doi [\apj] {10.1088/0004-637X/744/2/103}, \href
  {http://adsabs.harvard.edu/abs/2012ApJ...744..103M} {744, 103}

\bibitem[\protect\citeauthoryear{{Moreno M{\'e}ndez}, {L{\'o}pez-C{\'a}mara}
  \& {De Colle}}{{Moreno M{\'e}ndez} et~al.}{2017}]{Morenomendez+17}
{Moreno M{\'e}ndez} E.,  {L{\'o}pez-C{\'a}mara} D.,   {De Colle} F.,  2017,
  \mn@doi [\mnras] {10.1093/mnras/stx1385}, \href
  {http://adsabs.harvard.edu/abs/2017MNRAS.470.2929M} {470, 2929}

\bibitem[\protect\citeauthoryear{{Murguia-Berthier}, {MacLeod}, {Ramirez-Ruiz},
  {Antoni}  \& {Macias}}{{Murguia-Berthier} et~al.}{2017}]{Murguia-berthier+17}
{Murguia-Berthier} A.,  {MacLeod} M.,  {Ramirez-Ruiz} E.,  {Antoni} A.,
  {Macias} P.,  2017, \mn@doi [\apj] {10.3847/1538-4357/aa8140}, \href
  {http://adsabs.harvard.edu/abs/2017ApJ...845..173M} {845, 173}

\bibitem[\protect\citeauthoryear{{Nandez}, {Ivanova}  \& {Lombardi}}{{Nandez}
  et~al.}{2014}]{Nandez+14}
{Nandez} J.~L.~A.,  {Ivanova} N.,   {Lombardi} Jr. J.~C.,  2014, \mn@doi [\apj]
  {10.1088/0004-637X/786/1/39}, \href
  {http://adsabs.harvard.edu/abs/2014ApJ...786...39N} {786, 39}

\bibitem[\protect\citeauthoryear{{Nandez}, {Ivanova}  \& {Lombardi}}{{Nandez}
  et~al.}{2015}]{Nandez+15}
{Nandez} J.~L.~A.,  {Ivanova} N.,   {Lombardi} J.~C.,  2015, \mn@doi [\mnras]
  {10.1093/mnrasl/slv043}, \href
  {http://adsabs.harvard.edu/abs/2015MNRAS.450L..39N} {450, L39}

\bibitem[\protect\citeauthoryear{{Nelemans}, {Verbunt}, {Yungelson}  \&
  {Portegies Zwart}}{{Nelemans} et~al.}{2000}]{Nelemans+00}
{Nelemans} G.,  {Verbunt} F.,  {Yungelson} L.~R.,   {Portegies Zwart} S.~F.,
  2000, \aap, \href {http://adsabs.harvard.edu/abs/2000A%26A...360.1011N} {360,
  1011}

\bibitem[\protect\citeauthoryear{{Nordhaus} \& {Blackman}}{{Nordhaus} \&
  {Blackman}}{2006}]{Nordhaus+Blackman06}
{Nordhaus} J.,  {Blackman} E.~G.,  2006, \mn@doi [\mnras]
  {10.1111/j.1365-2966.2006.10625.x}, \href
  {http://adsabs.harvard.edu/abs/2006MNRAS.370.2004N} {370, 2004}

\bibitem[\protect\citeauthoryear{{Nordhaus}, {Blackman}  \& {Frank}}{{Nordhaus}
  et~al.}{2007}]{Nordhaus+07}
{Nordhaus} J.,  {Blackman} E.~G.,   {Frank} A.,  2007, \mn@doi [\mnras]
  {10.1111/j.1365-2966.2007.11417.x}, \href
  {http://adsabs.harvard.edu/abs/2007MNRAS.376..599N} {376, 599}

\bibitem[\protect\citeauthoryear{{Nordhaus}, {Wellons}, {Spiegel}, {Metzger}
  \& {Blackman}}{{Nordhaus} et~al.}{2011}]{Nordhaus+11}
{Nordhaus} J.,  {Wellons} S.,  {Spiegel} D.~S.,  {Metzger} B.~D.,   {Blackman}
  E.~G.,  2011, \mn@doi [Proceedings of the National Academy of Science]
  {10.1073/pnas.1015005108}, \href
  {http://adsabs.harvard.edu/abs/2011PNAS..108.3135N} {108, 3135}

\bibitem[\protect\citeauthoryear{{Ohlmann}, {R{\"o}pke}, {Pakmor}  \&
  {Springel}}{{Ohlmann} et~al.}{2016}]{Ohlmann+16a}
{Ohlmann} S.~T.,  {R{\"o}pke} F.~K.,  {Pakmor} R.,   {Springel} V.,  2016,
  \mn@doi [\apjl] {10.3847/2041-8205/816/1/L9}, \href
  {http://adsabs.harvard.edu/abs/2016ApJ...816L...9O} {816, L9}

\bibitem[\protect\citeauthoryear{{Ohlmann}, {R{\"o}pke}, {Pakmor}  \&
  {Springel}}{{Ohlmann} et~al.}{2017}]{Ohlmann+17}
{Ohlmann} S.~T.,  {R{\"o}pke} F.~K.,  {Pakmor} R.,   {Springel} V.,  2017,
  \mn@doi [\aap] {10.1051/0004-6361/201629692}, \href
  {http://adsabs.harvard.edu/abs/2017A%26A...599A...5O} {599, A5}

\bibitem[\protect\citeauthoryear{{Paczynski}}{{Paczynski}}{1976}]{Paczynski76}
{Paczynski} B.,  1976, in {Eggleton} P.,  {Mitton} S.,   {Whelan} J.,  eds,
  IAU Symposium Vol. 73, Structure and Evolution of Close Binary Systems. p.~75

\bibitem[\protect\citeauthoryear{{Passy} et~al.,}{{Passy}
  et~al.}{2012}]{Passy+12a}
{Passy} J.-C.,  et~al., 2012, \mn@doi [\apj] {10.1088/0004-637X/744/1/52},
  \href {http://adsabs.harvard.edu/abs/2012ApJ...744...52P} {744, 52}

\bibitem[\protect\citeauthoryear{{Paxton}, {Bildsten}, {Dotter}, {Herwig},
  {Lesaffre}  \& {Timmes}}{{Paxton} et~al.}{2011}]{Paxton+11}
{Paxton} B.,  {Bildsten} L.,  {Dotter} A.,  {Herwig} F.,  {Lesaffre} P.,
  {Timmes} F.,  2011, \mn@doi [\apjs] {10.1088/0067-0049/192/1/3}, \href
  {http://adsabs.harvard.edu/abs/2011ApJS..192....3P} {192, 3}

\bibitem[\protect\citeauthoryear{{Paxton} et~al.,}{{Paxton}
  et~al.}{2013}]{Paxton+13}
{Paxton} B.,  et~al., 2013, \mn@doi [\apjs] {10.1088/0067-0049/208/1/4}, \href
  {http://adsabs.harvard.edu/abs/2013ApJS..208....4P} {208, 4}

\bibitem[\protect\citeauthoryear{{Paxton} et~al.,}{{Paxton}
  et~al.}{2015}]{Paxton+15}
{Paxton} B.,  et~al., 2015, \mn@doi [\apjs] {10.1088/0067-0049/220/1/15}, \href
  {http://adsabs.harvard.edu/abs/2015ApJS..220...15P} {220, 15}

\bibitem[\protect\citeauthoryear{{Rasio} \& {Livio}}{{Rasio} \&
  {Livio}}{1996}]{Rasio+Livio96}
{Rasio} F.~A.,  {Livio} M.,  1996, \mn@doi [\apj] {10.1086/177975}, \href
  {http://adsabs.harvard.edu/abs/1996ApJ...471..366R} {471, 366}

\bibitem[\protect\citeauthoryear{{Reyes-Ruiz} \& {L{\'o}pez}}{{Reyes-Ruiz} \&
  {L{\'o}pez}}{1999}]{Reyes-Ruiz+1999}
{Reyes-Ruiz} M.,  {L{\'o}pez} J.~A.,  1999, \mn@doi [\apj] {10.1086/307827},
  \href {http://adsabs.harvard.edu/abs/1999ApJ...524..952R} {524, 952}

\bibitem[\protect\citeauthoryear{{Ricker} \& {Taam}}{{Ricker} \&
  {Taam}}{2008}]{Ricker+Taam08}
{Ricker} P.~M.,  {Taam} R.~E.,  2008, \mn@doi [\apjl] {10.1086/526343}, \href
  {http://adsabs.harvard.edu/abs/2008ApJ...672L..41R} {672, L41}

\bibitem[\protect\citeauthoryear{{Ricker} \& {Taam}}{{Ricker} \&
  {Taam}}{2012}]{Ricker+Taam12}
{Ricker} P.~M.,  {Taam} R.~E.,  2012, \mn@doi [\apj]
  {10.1088/0004-637X/746/1/74}, \href
  {http://adsabs.harvard.edu/abs/2012ApJ...746...74R} {746, 74}

\bibitem[\protect\citeauthoryear{{Sabach} \& {Soker}}{{Sabach} \&
  {Soker}}{2015}]{Sabach+Soker15}
{Sabach} E.,  {Soker} N.,  2015, \mn@doi [\mnras] {10.1093/mnras/stv717}, \href
  {http://adsabs.harvard.edu/abs/2015MNRAS.450.1716S} {450, 1716}

\bibitem[\protect\citeauthoryear{{Sabach}, {Hillel}, {Schreier}  \&
  {Soker}}{{Sabach} et~al.}{2017}]{Sabach+17}
{Sabach} E.,  {Hillel} S.,  {Schreier} R.,   {Soker} N.,  2017, \mn@doi
  [\mnras] {10.1093/mnras/stx2272}, \href
  {http://adsabs.harvard.edu/abs/2017MNRAS.472.4361S} {472, 4361}

\bibitem[\protect\citeauthoryear{{Sahai}, {Vlemmings}, {Gledhill}, {S{\'a}nchez
  Contreras}, {Lagadec}, {Nyman}  \& {Quintana-Lacaci}}{{Sahai}
  et~al.}{2017}]{Sahai+2017}
{Sahai} R.,  {Vlemmings} W.~H.~T.,  {Gledhill} T.,  {S{\'a}nchez Contreras} C.,
   {Lagadec} E.,  {Nyman} L.-{\AA}.,   {Quintana-Lacaci} G.,  2017, \mn@doi
  [\apjl] {10.3847/2041-8213/835/1/L13}, \href
  {http://adsabs.harvard.edu/abs/2017ApJ...835L..13S} {835, L13}

\bibitem[\protect\citeauthoryear{{Sandquist}, {Taam}, {Chen}, {Bodenheimer}  \&
  {Burkert}}{{Sandquist} et~al.}{1998}]{Sandquist+98}
{Sandquist} E.~L.,  {Taam} R.~E.,  {Chen} X.,  {Bodenheimer} P.,   {Burkert}
  A.,  1998, \mn@doi [\apj] {10.1086/305778}, \href
  {http://adsabs.harvard.edu/abs/1998ApJ...500..909S} {500, 909}

\bibitem[\protect\citeauthoryear{{Sandquist}, {Taam}  \& {Burkert}}{{Sandquist}
  et~al.}{2000}]{Sandquist+00}
{Sandquist} E.~L.,  {Taam} R.~E.,   {Burkert} A.,  2000, \mn@doi [\apj]
  {10.1086/308687}, \href {http://adsabs.harvard.edu/abs/2000ApJ...533..984S}
  {533, 984}

\bibitem[\protect\citeauthoryear{{Shiber} \& {Soker}}{{Shiber} \&
  {Soker}}{2018}]{Shiber+Soker18}
{Shiber} S.,  {Soker} N.,  2018, \mn@doi [\mnras] {10.1093/mnras/sty843}, \href
  {http://adsabs.harvard.edu/abs/2018MNRAS.tmp..819S} {}

\bibitem[\protect\citeauthoryear{{Shiber}, {Schreier}  \& {Soker}}{{Shiber}
  et~al.}{2016}]{Shiber+16}
{Shiber} S.,  {Schreier} R.,   {Soker} N.,  2016, \mn@doi [Research in
  Astronomy and Astrophysics] {10.1088/1674-4527/16/7/117}, \href
  {http://adsabs.harvard.edu/abs/2016RAA....16..117S} {16, 117}

\bibitem[\protect\citeauthoryear{{Shiber}, {Kashi}  \& {Soker}}{{Shiber}
  et~al.}{2017}]{Shiber+17}
{Shiber} S.,  {Kashi} A.,   {Soker} N.,  2017, \mn@doi [\mnras]
  {10.1093/mnrasl/slw208}, \href
  {http://adsabs.harvard.edu/abs/2017MNRAS.465L..54S} {465, L54}

\bibitem[\protect\citeauthoryear{{Soker}}{{Soker}}{1994}]{Soker94}
{Soker} N.,  1994, \mn@doi [\mnras] {10.1093/mnras/270.4.774}, \href
  {http://adsabs.harvard.edu/abs/1994MNRAS.270..774S} {270, 774}

\bibitem[\protect\citeauthoryear{{Soker}}{{Soker}}{2015}]{Soker15}
{Soker} N.,  2015, \mn@doi [\apj] {10.1088/0004-637X/800/2/114}, \href
  {http://adsabs.harvard.edu/abs/2015ApJ...800..114S} {800, 114}

\bibitem[\protect\citeauthoryear{{Soker}}{{Soker}}{2017}]{Soker17b}
{Soker} N.,  2017, \mn@doi [\mnras] {10.1093/mnras/stx1978}, \href
  {http://adsabs.harvard.edu/abs/2017MNRAS.471.4839S} {471, 4839}

\bibitem[\protect\citeauthoryear{{Soker} \& {Harpaz}}{{Soker} \&
  {Harpaz}}{2003}]{Soker+Harpaz03}
{Soker} N.,  {Harpaz} A.,  2003, \mn@doi [\mnras]
  {10.1046/j.1365-8711.2003.06689.x}, \href
  {http://adsabs.harvard.edu/abs/2003MNRAS.343..456S} {343, 456}

\bibitem[\protect\citeauthoryear{{Soker} \& {Rappaport}}{{Soker} \&
  {Rappaport}}{2000}]{Soker+2000}
{Soker} N.,  {Rappaport} S.,  2000, \mn@doi [\apj] {10.1086/309112}, \href
  {http://adsabs.harvard.edu/abs/2000ApJ...538..241S} {538, 241}

\bibitem[\protect\citeauthoryear{{Soker} \& {Rappaport}}{{Soker} \&
  {Rappaport}}{2001}]{Soker+2001}
{Soker} N.,  {Rappaport} S.,  2001, \mn@doi [\apj] {10.1086/321669}, \href
  {http://adsabs.harvard.edu/abs/2001ApJ...557..256S} {557, 256}

\bibitem[\protect\citeauthoryear{{Springel}}{{Springel}}{2010}]{Springel10}
{Springel} V.,  2010, \mn@doi [\mnras] {10.1111/j.1365-2966.2009.15715.x},
  \href {http://adsabs.harvard.edu/abs/2010MNRAS.401..791S} {401, 791}

\bibitem[\protect\citeauthoryear{{Staff}, {De Marco}, {Macdonald}, {Galaviz},
  {Passy}, {Iaconi}  \& {Low}}{{Staff} et~al.}{2016}]{Staff+16}
{Staff} J.~E.,  {De Marco} O.,  {Macdonald} D.,  {Galaviz} P.,  {Passy} J.-C.,
  {Iaconi} R.,   {Low} M.-M.~M.,  2016, \mn@doi [\mnras]
  {10.1093/mnras/stv2548}, \href
  {http://adsabs.harvard.edu/abs/2016MNRAS.455.3511S} {455, 3511}

\bibitem[\protect\citeauthoryear{{Webbink}}{{Webbink}}{2008}]{Webbink08}
{Webbink} R.~F.,  2008, in {Milone} E.~F.,  {Leahy} D.~A.,   {Hobill} D.~W.,
  eds,  Astrophysics and Space Science Library Vol. 352, Astrophysics and Space
  Science Library. p.~233 (\mn@eprint {arXiv} {0704.0280}),
  \mn@doi{10.1007/978-1-4020-6544-6_13}

\bibitem[\protect\citeauthoryear{{Witt}, {Vijh}, {Hobbs}, {Aufdenberg},
  {Thorburn}  \& {York}}{{Witt} et~al.}{2009}]{Witt+2009}
{Witt} A.~N.,  {Vijh} U.~P.,  {Hobbs} L.~M.,  {Aufdenberg} J.~P.,  {Thorburn}
  J.~A.,   {York} D.~G.,  2009, \mn@doi [\apj] {10.1088/0004-637X/693/2/1946},
  \href {http://adsabs.harvard.edu/abs/2009ApJ...693.1946W} {693, 1946}

\bibitem[\protect\citeauthoryear{{de Kool}}{{de Kool}}{1990}]{Dekool90}
{de Kool} M.,  1990, \mn@doi [\apj] {10.1086/168974}, \href
  {http://adsabs.harvard.edu/abs/1990ApJ...358..189D} {358, 189}

\makeatother
\end{thebibliography}
}

\appendix

\section{Lower resolution tests to explore dependence on numerical parameters}
\label{sec:test_runs}
In this section we compare results of Model~B with five other runs that were done in the testing phase of the project
and, like Model~B, include Krumholz subgrid accretion.
The setups are all similar to that of Model~B, 
with small differences in the initial conditions and numerical parameters, especially involving resolution.
Although it was not feasible to perform a carefully controlled convergence study with current resources,
comparing these six simulations can give us a sense of any dependence of the results on numerical parameters.

First we briefly describe each test run.
Model~B has been described in detail in Section~\ref{sec:methods},
and here we mention the differences in the parameters of each test run as compared to those of Model~B.
Some parameters, notably box size, boundary conditions, accretion radius measured in grid cells (=4), 
and pre-$t=0$ relaxation (velocity damping) procedure, are the same for all of these runs.%
\footnote{It is worth noting that the relaxation runs differ from each other 
only in that the softening length in the relaxation run is the same as that in the respective binary run,
and the resolution of the relaxation run is the same as the initial resolution of the binary run.
Model~A does not employ a relaxation run.}
In all of the test runs, the highest resolution is a factor of two lower than in Model~B,
with a smallest resolution element of $0.29R_\odot$ instead of $0.14R_\odot$.
The softening length in the test runs is equal to that in Model~B, except for in Model~C,
where the softening length is twice as large, being equal to $4.8R_\odot$ instead of $2.4R_\odot$
The base resolution element in all of the test runs has size $9.0R_\odot$, which is four times larger than in Model~B.
The ambient pressure is the same in all the test runs, and is equal to $10^6\dynecmcm$, that is, one order of magnitude higher than in Model~B.
The ambient density is set to $6.7\times10^{-9}\gcmcmcm$, the same as in Model~B, 
in all test runs except Model~G, where it is instead set to $10^{-10}\gcmcmcm$.
These properties are summarized in Table~\ref{tab:models}, where for completeness we have also included Model~A (no subgrid accretion).

Where each model differs more substantially is in the refinement algorithm.
In Models~C and E, the refinement is controlled internally by AstroBEAR,
and refines when one of three criteria---the density gradient, momentum gradient, or the potential due to gas self-gravity---exceeds some (default) threshold.%
\footnote{In practice the default algorithm tends to refine more extensively than desired for this particular application
(e.g. spiral shocks are highly resolved even very far away from the point particles).}
However, in the relaxation runs, we (in addition) force the refinement to be at the highest level inside a spherical region that contains the entire primary star.
Models~D and F are similar to Model~E with the only differences being the regions chosen for maximum refinement. 
In Model~D, the refinement is set to be at the highest level inside a sphere of radius $50R_\odot$ centred on the primary point particle,
as well as in a cylinder with axis parallel to $z$ centred on the secondary with radius and height equal to $50R_\odot$.
In Model~F the highest AMR level is activated within a sphere centred on the primary particle 
with radius equal to $\min(72R_\odot,1.5a)$, with $a$ equal to the inter-particle separation distance.
Model~G is identical to Model~E except that the ambient density is $10^{-10}\gcmcmcm$.
Finally, the sizes of buffer zones separating meshes of different refinement level also vary between models, 
with Model~D having no buffer zones and Models~B, C, E, F and G having buffer zones of 2 cells per level.
Model~A has buffer zones of 16 cells per level.

\begin{table*}
  \begin{center}
  \caption{Parameters of the various models. Models~A and B are the main ones studied in this work while Models~C through G are for testing.
           Parameters are the base resolution element $\delta\base$, highest refinement level resolution element $\delta\ma$,
           side length of the full domain $L$, spline softening length $r\soft$, ambient density $\rho\amb$, ambient pressure $P\amb$,
           and accretion radius in the Krumholz subgrid model (measured in grid cells).           
           Models also differ in the way that the refinement is carried out (see text).
           Model~A is included for completeness, but results are not compared with those of Models~C through G 
           because Model~A does not have a subgrid accretion model implemented.
           For Model~A, an arrow signifies that the quantity is modified at the time $t=16.7\da$.
           \label{tab:models}
          }
  \begin{tabular}{@{}lccccccc@{}}
    \hline
    Model  &$\delta\base$ &$\delta\ma$           &$L$          &$r\soft$             &$\rho\amb$           &$P\amb$           &Accretion radius\\
           &$[R_\odot]$   &$[R_\odot]$           &$[R_\odot]$  &$[R_\odot]$          &$[10^{-10}\gcmcmcm]$ &$[10^5\dynecmcm]$ &$[\delta\ma]$\\
    \hline                   
    A      &$2.2$         &$0.14\rightarrow0.07$ &$1150$       &$2.4\rightarrow1.2$  &$67$                 &$1$               &---\\
    B      &$2.2$         &$0.14$                &$575$        &$2.4$                &$67$                 &$1$               &4\\
    C      &$9.0$         &$0.29$                &$575$        &$4.8$                &$67$                 &$10$              &4\\
    D      &$9.0$         &$0.29$                &$575$        &$2.4$                &$67$                 &$10$              &4\\
    E      &$9.0$         &$0.29$                &$575$        &$2.4$                &$67$                 &$10$              &4\\
    F      &$9.0$         &$0.29$                &$575$        &$2.4$                &$67$                 &$10$              &4\\
    G      &$9.0$         &$0.29$                &$575$        &$2.4$                &$1$                  &$10$              &4\\
    \hline
  \end{tabular}
  \end{center}
\end{table*}

\begin{figure*}
  \includegraphics[width=\textwidth,clip=true,trim= 0 0 0 0]{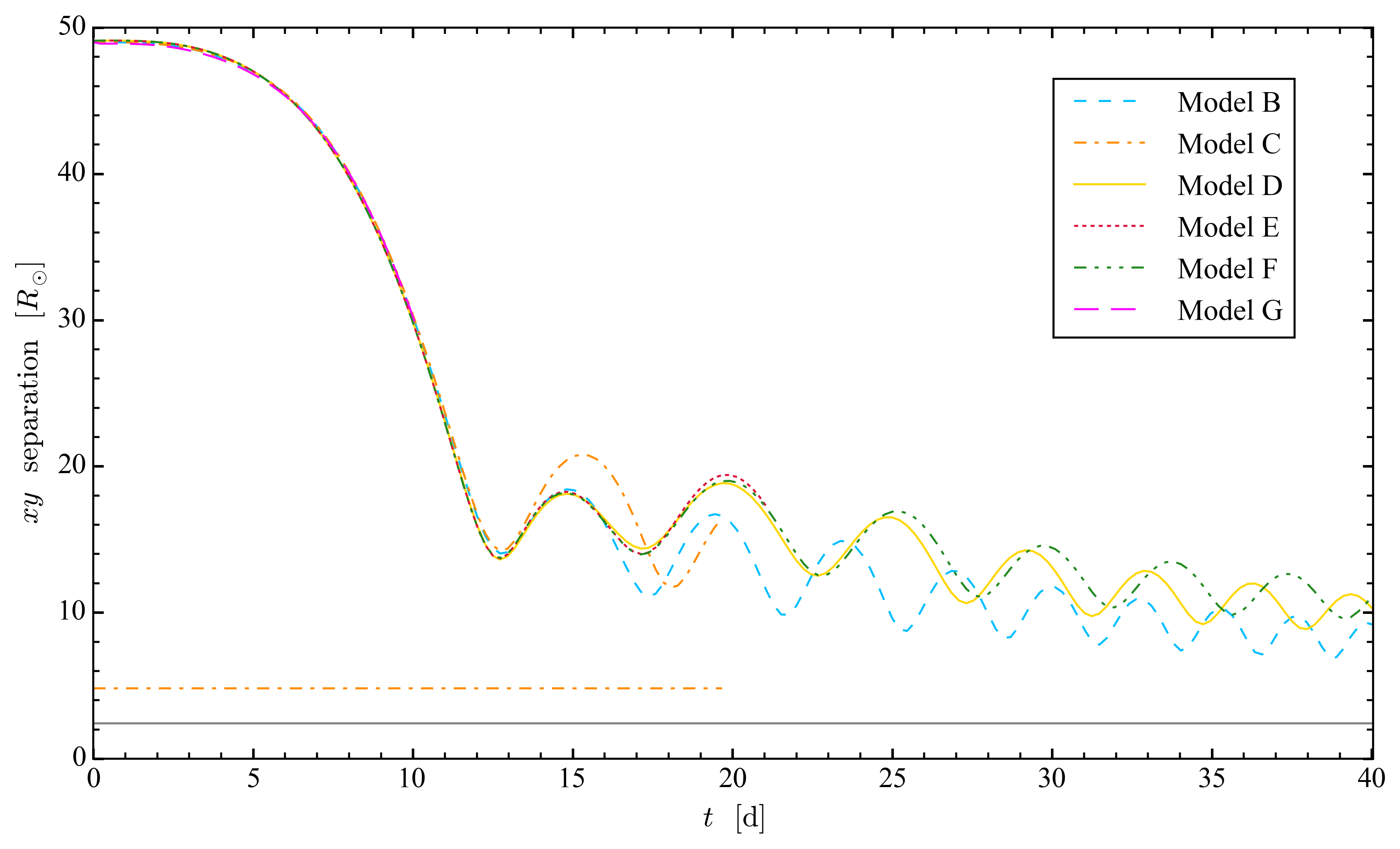} 
  \caption{Similar to Figure~\ref{fig:orbit} but now showing inter-particle separation in the orbital plane for Models~B through G.
           Horizontal lines show the spline softening length for Model~C (orange dash-dotted) and all other models (grey solid).
           \label{fig:orbit_test}
          }
\end{figure*}

\begin{figure*}
  \includegraphics[width=\textwidth,clip=true,trim= 0 0 0 0]{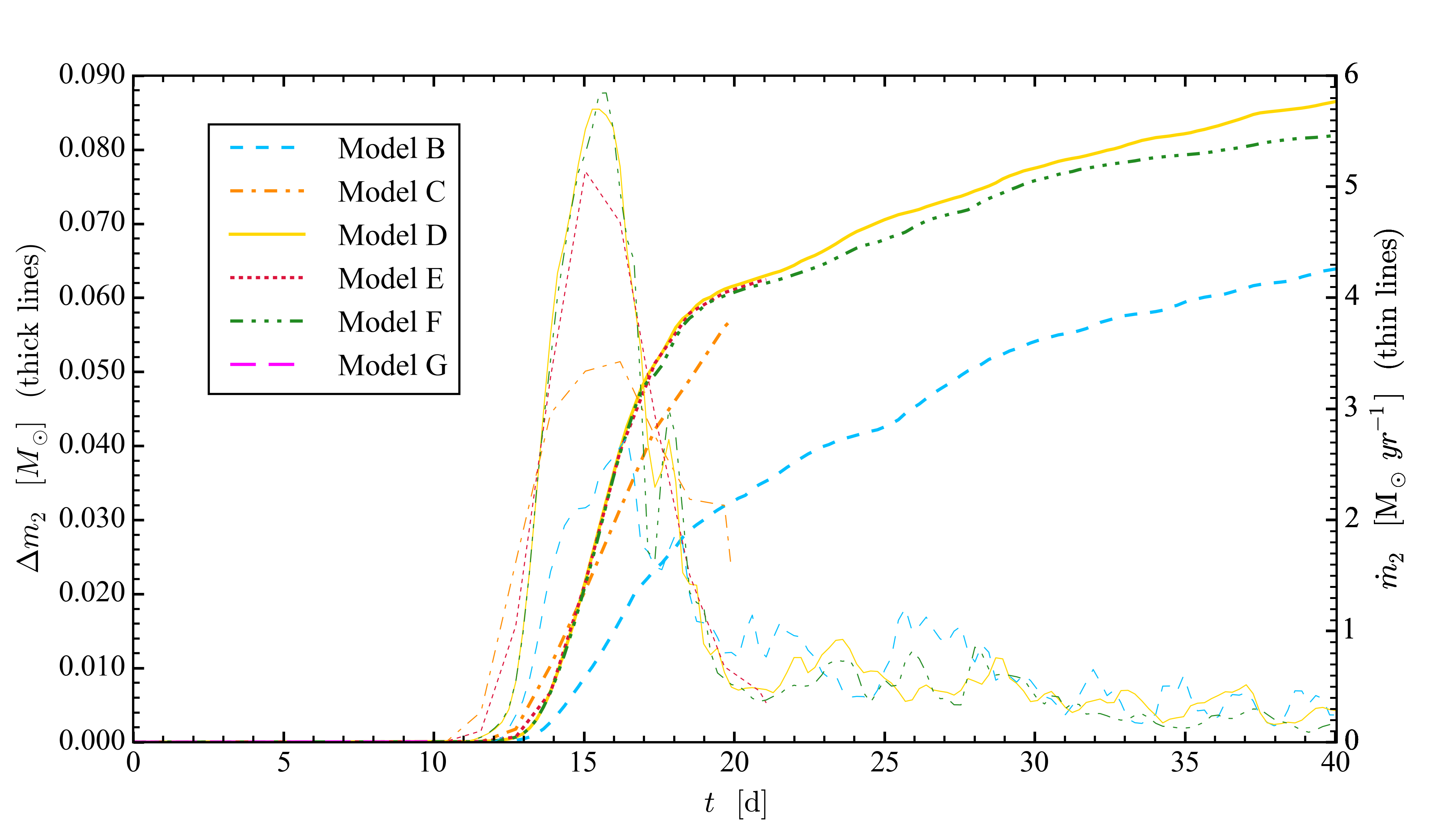}
  \caption{Similar to the lower panel of Figure~\ref{fig:macc} for Model~B, 
           but now showing \citetalias{Krumholz+04} accretion rates onto the secondary for Models~B through G.
           The accreted mass is shown by thick lines while its rate is shown by thin lines.
           (The sampling rate is smaller for some of the test runs because not all of the original data was retained.)
           \label{fig:macc_test}
          }
\end{figure*}

Results for the evolution of the inter-particle separation for Models~B through G are shown in Figure~\ref{fig:orbit_test}.
The softening length is drawn as a horizontal line, in solid black for all models except Model~C, which is shown orange dash-dotted.
We see that the separation at first apastron passage is larger for Model~C than for the other models.
The larger softening length has affected the accumulation of gas around the particles, 
leading to a reduction in the gravitational force.
This also leads to a reduction in the maximum accretion rate, as shown in Figure~\ref{fig:macc_test},
where we plot the accreted mass (thick lines) and accretion rate (thin lines) for the various models.
Models~D through G differ from Model~B mainly in the resolution, with smallest resolution element larger by a factor of two compared to Model~B.
We see that this leads to an important difference in the orbit for Models~D through F (Model~G is not evolved this far in time), 
but only after the first apastron passage.
The maximum accretion rate is much larger for Models~D through F than for Model~B.
This is likely caused by the fact that the accretion radius, which is fixed to 4 grid cells, 
is actually twice as large in physical units, because of the lower resolution, 
so that accretion is allowed to take place within a larger volume around the secondary.
However, it is not clear why this causes a larger separation at the second periastron passage.
Further, we can infer from Figures~\ref{fig:orbit_test} and \ref{fig:macc_test} that differences in the choices of refinement region in Models~D, E and F
have a relatively minor effect on the orbital evolution and accretion rate.
However, by the fourth apastron passage at $t\approx25\da$ there are significant differences between Models~D and F.
The refinement algorithm in Model~F is probably too aggressive because the orbital frequency is smaller than that of Model~D.

As noted above, the maximum accretion rates vary between runs, 
and this is likely due to the differences in softening length and accretion radius (in physical units).
However, it is reassuring that the accretion rates between $t\approx18\da$ and $t=40\da$ 
are comparable for Models~B, D, E (which extends only up to $t\approx21\da$) and F.
We can conclude that our estimates of the accretion rate after the initial burst of accretion are not particularly sensitive to resolution and accretion radius.

\end{document}